\documentclass[twocolumn]{aa}
\usepackage{graphicx}
\usepackage[varg]{txfonts}
\usepackage{enumitem}
\usepackage{natbib}   
\bibpunct{(}{)}{;}{a}{}{,}  
\usepackage{amsmath,amssymb}
\usepackage{mathtools}
\usepackage{lscape}
\usepackage{longtable}
\usepackage{enumerate}
\usepackage{setspace}
\usepackage[usenames,dvipsnames]{color}
\usepackage[draft, pagebackref=true]{hyperref}
\hypersetup{colorlinks=true,linkcolor=blue,citecolor=blue,filecolor=magenta,urlcolor=blue}
\usepackage{float,morefloats} 
\begin{document}

\title{The impact of the metallicity and star formation rate on the time-dependent galaxy-wide 
stellar initial mass function}
\author{T. Jerabkova \inst{1,2,3} \fnmsep\thanks{tereza.jerabkova@eso.org}
\and A. H. Zonoozi \inst{2,4}
\and P. Kroupa \inst{2,3}
\and G. Beccari \inst{1}
\and Z. Yan \inst{2,3}
\and A. Vazdekis \inst{5}
\and Z.-Y. Zhang \inst{6,1}
}
\institute{
European Southern Observatory, Karl-Schwarzschild-Straße 2, 85748 Garching bei München	
\and
Helmholtz Institut f\"{u}r Strahlen und Kernphysik, Universit\"{a}t Bonn, Nussallee 14–16, 53115 Bonn, Germany
\and
Astronomical Institute, Charles University in Prague, V 
Hole\v{s}ovi\v{c}k\'ach 2, CZ-180 00 Praha 8, Czech Republic
\and
Department of Physics, Institute for Advanced Studies in Basic Sciences (IASBS), PO Box 11365-9161, Zanjan, Iran
\and
Instituto de Astrofisica de Canarias,E-38200 La Laguna, Tenerife, Spain
\and 
Institute for Astronomy, University of Edinburgh, Blackford Hill,
Edinburgh, EH9 3HJ, UK
}
\date{Received 20.03.2018; accepted 24.08.2018}
\titlerunning{The impact of metallicity and star formation rate on the galaxy-wide 
initial mass function}

\abstract{
The stellar initial mass function (IMF) is commonly assumed to be an invariant
probability density distribution function of initial stellar masses, being generally represented by the canonical IMF which we define to be the result of one star formation event in an embedded cluster. As a consequence the galaxy-wide IMF (gwIMF), defined as the sum of the IMFs of all star forming regions in which embedded clusters form which spawn the galactic field population of the galaxy, should also be invariant and of the same form as the canonical IMF. 
Recent observational and theoretical results challenge the hypothesis that the gwIMF is invariant. In order to study the possible reasons for this variation, it is useful to relate the observed IMF to the gwIMF. Starting with the IMF determined in resolved star clusters, we apply the IGIMF-theory to calculate a comprehensive grid of gwIMF models  for  metallicities, $\mathrm{[Fe/H]}\in (-3,1)$ and galaxy-wide star formation rates, $\mathrm{SFR}\in(10^{-5},10^{5})\,\mathrm{M_{\odot}/yr}$. For a galaxy with metallicy [Fe/H]$<0$ and SFR$\,> 1\,M_\odot$/yr, which is a common condition in the early Universe,  we find that the gwIMF is both bottom-light (relatively fewer low-mass stars) and top-heavy (more massive stars), when compared to the canonical IMF. 
For a SFR $< 1\,\mathrm{M_{\odot}/yr}$ the gwIMF becomes top-light regardless of the metallicity. For metallicities $\mathrm{[Fe/H]} > 0$ the gwIMF can become bottom-heavy regardless of the SFR. 
The IGIMF models predict that massive elliptical galaxies should have formed with a gwIMF that is top-heavy within the first few hundred Myr of the galaxy's life and that it evolves into a bottom-heavy gwIMF in the metal-enriched galactic center. 
Using the gwIMF grids, we study the SFR$-$H$\alpha$ relation, its dependency on metallicity and the SFR, the correction factors to the Kennicutt SFR$_{\rm K}-$H$\alpha$ relation, and provide new fitting functions. 
Late-type dwarf galaxies show significantly higher SFRs with respect to Kennicutt SFRs, while star forming massive galaxies have significantly lower SFRs than hitherto thought. This has implications for the gas-consumption time scales and for the main sequence of galaxies. The Leo P and ultra-faint dwarf galaxies are discussed explicitly.
}
 \maketitle

\section{Introduction}
\label{sec:Introd}
The stellar initial mass function (IMF) is a theoretical representation of the number distribution of stellar masses at their birth formed in one star-formation event.
The IMF is often described as  $\xi_{\star}(m) = {\rm d}N/{\rm d}m$, where d$N$ is the number of stars formed locally\footnote{Note that "local" is used here to mean a small region in a galaxy and in this case an embedded-cluster-forming molecular cloud core. It is not the Solar neighborhood.} in the mass interval $m$ to $m+${\rm d}$m$.
The IMF can be conveniently mathematically 
expressed in the form of a multi-power law with indices $\alpha^{\mathrm{can}}_1 \approx 1.3$ for stars in the mass range 
$0.1-0.5\,M_{\odot}$ and $\alpha^{\mathrm{can}}_2\approx 2.3$ for stars more massive than $0.5\,M_{\odot}$. That this canonical stellar IMF is an invariant probability density distribution function of stellar masses is usually considered to be a null hypothesis and a benchmark for stellar population studies \citep[e.g.][]{Selman2008, Kroupa2013, Offner2014, Jerabkova2017, KJ2018}.

The detailed form of the IMF is relevant for almost all fields related to star-formation, thus it has important implications for the luminous, dynamical and chemical evolution of stellar populations. 
In studies of both Galactic and extra-galactic integrated systems,  an IMF needs to be assumed to derive the star-formation rates (SFRs) by extrapolating from massive stars that always dominate the luminosities.  
The IMF is therefore a fundamental entity entering directly or indirectly into many astrophysical problems \citep[e.g.][]{Kroupa2002,Chabrier2003,Bastian2010, Kroupa2013, Offner2014, Hopkins2018}. Observational studies of nearby star forming regions suggest 
that stars form in dense cores inside molecular clouds 
\citep{Lada2003,Lada2010,Kirk2011,Gieles2012,Kirk2012,Megeath2016,Stephens2017,Ramirez+16,Hacar+17,Lucas2018} usually following the canonical stellar IMF \citep[e.g.][]{Kroupa2002,Chabrier2003,Bastian2010, Kroupa2013}.  

The canonical stellar IMF is derived from observations of field stars and nearby star forming regions that form stars in local over-densities called embedded star clusters or correlated star formation events (CSFEs) that are approximately $1\,\mathrm{pc}$ across and form a population of stars on a time scale of $1\,\mathrm{Myr}$ (\citealt{Kroupa2013, Yan2017} and references therein). The molecular clouds as a whole are not self-gravitationally bound in the majority of cases \citep{Hartmann2001,Elmegreen2002,Elmegreen2007,Ballesteros2007,Dobbs2011,Lim+18}. However, their complex filamentary substructures on sub-pc scales can be locally gravitationally bound and also gravitationally unstable \citep{Hacar+17,Hacar+17b},
which  may set the initial conditions for the formation of stars.
Star formation happens in correlated dense regions of molecular gas, which have intrinsic physical connections,  instead of being distributed everywhere randomly inside molecular clouds. For practical purposes we refer to these CSFEs as embedded clusters. 
For the computations of the gwIMF only the stellar census matters. Nevertheless for the interpretation of observations of stellar populations in galaxies the initial conditions in star forming regions are relevant. The CSFEs/embedded clusters are potentially dynamically very active \citep{KroupaBoily2002, Kroupa2005,Oh2015,Oh2016, Brinkmann2017}\footnote{The Nbody star cluster evolution computations available on youtube, "Dynamical ejection of massive stars from a young star cluster" by Seungkyung Oh based on \cite{Oh2015,Oh2016}, demonstrate how dynamically active the binary-rich very young clusters are in dispersing their stars to relatively large distances.} and thus very quickly, within fractions of a Myr,  they spread their stars out through the star forming regions and later to the galactic field. Therefore, the dynamical processes on a star cluster scale need to be taken into account to obtain a physically correct picture of young stellar populations and of their distribution.

Increasing observational evidence 
suggests that star formation is a self-regulated process rather than being a purely stochastic one \citep{Papadopoulos2010, Kroupa2013,Kroupa2015, Yan2017, Lim+18,Plunkett2018}.  That the star formation efficiency of embedded clusters  is always found, observationally and theoretically, to be less than 30-40~per cent \citep{AF96,Lada2003,Lada2010, Hansen2012,Machida2012, Federrath2014,Federrath2015,Megeath2016} supports self-regulation: the gravitational collapse of an embedded-cluster forming cloud core leads to star formation which heats, ionizies and removes gas from the core. This may be a reason why a most-massive-star--embedded-cluster-mass relation might exist \citep{Weidner2010}.

For nearby resolved star-forming regions, the IMF can be understood as describing a single star-formation event happening on a physical scale of about $1\,$pc; and beyond this scale the molecular gas is gravitationally unstable and would form individual embedded clusters or small groups of stars. However, it is non-trivial to calculate the IMF of an unresolved stellar population, for example, of a whole galaxy, because it contains many stellar clusters formed at different times. 
The galaxy wide IMF (gwIMF) is, on the other hand, the sum of all the IMFs of all star forming regions belonging to a given galaxy \citep[e.g.][see also Fig.~\ref{fig:gal_sk}]{KroupaWeidner2003, Weidner2013,Kroupa2013,Yan2017}. 
Assuming that the canonical IMF is a universal probability density distribution function, the shape of the gwIMF should be equal to that of the canonical IMF. On the other hand, if the gwIMF differs from the canonical IMF, then the canonical IMF cannot be universal and/or it cannot be described as a stationary probability density distribution function.

Therefore, a  fundamental question arises naturally: is the stellar IMF a universal probability density distribution function \citep{Kroupa2013}? An overabundance 
of low-mass stars ($<1\,M_{\odot}$) with respect to the canonical stellar IMF is called a \textit{bottom-heavy} IMF, and a deficit of low mass stars is a \textit{bottom-light} IMF. For the massive stars ($>1\,M_{\odot}$), an overabundance or deficit of stars relative to the canonical IMF results in a \textit{top-heavy} and a \textit{top-light} IMF,  respectively. Studies of globular clusters, ultra compact dwarf galaxies and young massive clusters suggest that in a low metallicity and high gas-density environment the stellar IMF may become top-heavy \citep[e.g][]{Dabringhausen2009,Dabringhausen2012,Marks2012TH,Zonoozi2016,Haghi2017,Kalari2018,Schneider2018} while being bottom-heavy in metal rich ($Z>Z_\odot$) environments \citep{Kroupa2002,Marks2012}. 
Such bottom heavy-IMFs have been found in the centers of nearby elliptical galaxies,  where the metallicities are higher than $Z_\odot$  \citep{vanDokkum2010,Conroy2017}.
The progenitors of elliptical galaxies, on the other hand, are suggested to have had top-heavy gwIMFs instead, based on the evolution of their chemical composition \citep{Matteucci1994,Vazdekis1997, Weidner2013,Ferreras+15, MartinNavarro16}.
Top-heavy gwIMFs are often found in galaxies with high star-formation rates (SFRs; \citealt{Gunawardhana2011,Fontanot2017,DeMasi2018a,Zhang2018,Fontanot2018a,Fontanot2018b}), while top-light gwIMFs are evident in galaxies with low SFRs (e.g. \citealt{Lee2009,Meurer2009,Watts2018}).
Interesting in this context is the newly introduced method of tracing the variation of the gwIMF using observations of CNO isotopes 
in the molecular ISM with ALMA, and it shows highly consistent results with the gwIMF theory \citep{Papadopoulos2010,Romano2017,Zhang2018}.
All this work suggests that the gwIMF is not in a constant form and that it deviates from the canonical IMF, depending on the star-formation activity.  These new findings challenge the idea that the IMF is a universal probability density distribution function. 

Here we study the variation of the gwIMF using the integrated 
galaxy-wide IMF (IGIMF) theory \citep{KroupaWeidner2003,Kroupa2013,Yan2017}. In this theory, the model of the gwIMF, i.e. the IGIMF,  is
constructed by summing (i.e. integrating) the IMFs of all star formation events in 
the whole galaxy at a given time. This results in a dependency of the gwIMF on the galaxy-wide SFR and metallicity, and therefore also on the time. 

With this contribution we investigate the full range of IGIMF variation. The novel aspect here is the incorporation of the metallicity dependence of the IMF as deduced by \cite{Marks2012TH} based on a stellar-dynamical study of evolved globular clusters, which also took into account constraints from ultra-compact dwarf galaxies by \cite{Dabringhausen2009, Dabringhausen2010, Dabringhausen2012}. These constraints on how the IMF varies with local physical conditions are independent from any variation of the gwIMF deduced from observation. Thus, if the observed variation of the gwIMF can be accounted for with these IMF variations then this will play an important role in the convergence of our understanding of stellar populations over cosmic time. 

Section~2 defines and clarifies terminology used further on,  Section~3 explains the IGIMF theory and its implementations. Results-section~4 presents the parametrized grid of galaxy-wide IMFs and discusses the implications. The implications include the evolution of the gwIMF in elliptical galaxies,
quantifying the correction factors for H$\alpha$-based SFR estimators, the case of the Leo~P dwarf galaxy and its very low SFR and massive star population, the baryonic-Tully-Fisher relation and ultra-faint dwarf galaxy satellites. In Sec.~5 an additional discussion is provided and Sec.~6 contains the conclusions.

We emphasis that this is the first time that a full grid of IGIMFs is made available in dependency of the galaxy-wide SFR and the galaxy metallicity.

\section{Terminology}
In the manuscript we frequently use four acronyms referring 
to the stellar initial mass function, i.e. IMF, cIMF, gwIMF and IGIMF. The IMF represents the stellar initial  mass function of stars formed during one star formation event in an 
initially gravitationally collapsing region in a molecular cloud (time scale $\approx 1$ Myr, spatial scale $\approx 1$ pc), that is in an embedded cluster. The cIMF (composite IMF) represents the sum of the IMFs over a larger region, such as
whole T-Tauri and OB associations or even a larger part of a galaxy.
The gwIMF is the initial stellar mass function of newly formed stars in a whole galaxy formed over a time scale $\delta t \approx 10$ Myr (see Sec.~\ref{sec:assumptions}) and can be inferred from observations or it can be computed. The IGIMF is the theoretical framework that allows us to compute the gwIMF.
For clarity the acronyms are summarized in Tab.~\ref{tab:acr}.
We use the term cIGIMF to refer to a theoretical formulation of the cIMF within the IGIMF framework. Once the region of interest is bigger than several molecular clouds (about $>100$ pc) the time scale  $\delta t$ would not change since the life-time of molecular clouds is about 10~Myr.

We emphasize that it is important to distinguish between the IMF, cIMF and gwIMF. The reason is that only if the star formation process is a stochastic invariant one (in the sense that once stars begin to form then the mass of the star is not being related to the local physical conditions) will the IMF be an invariant probability density distribution function. If however, the mass of the born star (which assembles to within about 95 per cent of its main sequence value within about $10^5\,$yr, \citealt{Wuchterl2003,Duarte13}) does depend on the local conditions, then the IMF will not be an invariant probability density distribution function: if the physical conditions in a embedded-cluster-forming molecular cloud core differ from those in another one, then the distribution of stellar masses will also differ. The recent ALMA observation of an extremely young embedded cluster shows the mm sources to be nearly perfectly mass-segregated suggesting that local physical conditions are indeed probably very important is determining which stars form \citep{Plunkett2018}.
A variation of the IMF with physical conditions has been expected from basic theory (see e.g. the discussion in \citealt{Kroupa2013}) but resolved observations of star forming regions in the Local Group have been indicating that the variations, if existing, are not detectable \citep{Kroupa2001,Kroupa2002,Bastian2010}. 

Thus, if the IMF is not an invariant probability density distribution function, then the sum of two star forming events will not be the same as one larger one with the same number of stars. The composite and galaxy-wide IMF will, in this case, differ from the IMF. An explicit observational example of this is reported for the Orion~A cloud by \cite{Hsu2012}. The physical and empirical evidence thus suggests that the gwIMF should be varying. The alternative, benchmark conservative model is to treat the IMF as a scale-free invariant probability density distribution function and to set gwIMF$=$stellar IMF taking into account the appropriate normalisation. This conservative hypothesis is referred to as the caninvgwIMF hypothesis according to which gwIMF is equal in form to the invariant canonical stellar IMF.

\begin{table}
\caption{The initial mass function acronyms summary.f}   
\label{tab:acr} 
\centering   
\begin{tabular}{c l} 
\hline\hline       
Acronym & explanation \\ 
\hline 
IMF & the stellar initial  mass function of stars formed \\
    &  during one star formation event in an initially \\
    &  gravitationally bound region \\
cIMF & composite-IMF, the sum of the IMFs over larger  \\
     & regions within a galaxy\\
gwIMF & the initial stellar mass function  of \textit{newly} \\
      & formed stars in a whole galaxy\\
IGIMF & the theoretical framework that allows us\\
      &   to compute the gwIMF\\
cIGIMF & cIMF computed withing IGIMF framework\\
caninvgwIMF & gwIMF is invariant and equal to the\\ 
       & invariant canonical stellar IMF\\
\hline                                             
\end{tabular}
\tablefoot{This table summarizes the acronyms and variables used in this paper to characterize initial stellar masses of stellar populations over different scales. As \textit{newly} we assume the time scale $\delta t = 10\,$Myr (Sec.~\ref{sec:assumptions}) in this text.  We note that the conservative bench-mark or null hypothesis to compare models with is the assumption that gwIMF$=$stellar IMF, assuming in this particular case that the stellar IMF is an invariant scale-free probability density distribution function. That is, in this case gwIMF is referred to as the canonical invariant gwIMF (caninvgwIMF) case.
Note also that "local" is used throughout this text to mean a small region in a galaxy. It is not the Solar neighborhood.}
\label{tab:tab1}
\end{table}

\section{Methods}

\begin{table}
\caption{Physical meanings of acronyms and model parameters.}              
\label{table:param}      
\centering                                      
\begin{tabular}{c c}          
\hline\hline       
Acronym & Physical meaning  \\    
\hline                                   
$m$ & integration ($\int$) variable for stellar mass \\      
$m_{\mathrm{max}}$ &  $\int$ upper limit for stellar mass\\
$m_{\mathrm{min}}$ & $\int$ lower limit for stellar mass\\
$m$  & stellar mass \\
$M$ & $\int$ variable for stellar mass of EC\\
$M_{\mathrm{max}}$ & $\int$ upper limit for stellar mass of EC mass \\
$M_{\mathrm{min}}$ & $\int$ lower limit for stellar mass of EC mass \\
$M_{\mathrm{ecl}}$ & stellar-mass of EC\\
$M_{\mathrm{ecl,max}}$ & most massive EC mass \\
$M_{\mathrm{cl}}$ & mass of the pre-cluster molecular cloud \\
SFR & galaxy-wide star formation rate\\
\hline                                             
\end{tabular}
\tablefoot{This table summarizes the used variables in this paper which are the same as used in \cite{Yan2017}. We note embedded cluster as EC in the table. }
\end{table}

The IGIMF theory is based on several assumptions which are described below in detail. We consider possible variations resulting in several different formulations that are all considered in this work. The assumptions, or axioms, are also detailed in \cite{Recchi2015}.

In a nutshell, the IGIMF theory spatially integrates over the whole galaxy by summing the local galactic 
star forming regions to obtain the gwIMF (of the newly formed stellar 
population) in a given time interval $\delta t$ (see Sec.~\ref{sec:assumptions}). 
Two approaches exist: Here (as well as in \citealt{Yan2017})  the first approach is used according to which the galaxy is treated as one unresolved object in which the integration over all freshly formed embedded clusters is performed without taking into account their spatial position and individual chemical properties. In this IGIMF approach the gwIMF is calculated at a particular time assuming all embedded clusters have the same metallicity. The second, spatially-resolved approach has also been pioneered \citep{Pflamm2008}, and in principle allows the embedded clusters to have different metallicities. 

In both approaches, 
the IMF in an individual embedded cluster follows the
empirical parametrization from mostly nearby (Galactic) observations of resolved stellar populations and varies with initial volume gas density of the embedded-cluster-forming cloud core or clump and its metallicity \citep{Marks2012TH,Marks2012}. 
The cosmological principle is assumed, in that the physical variations and associated IMF variations apply to the early Universe as well. That is, we assume that embedded clusters with the same mass, metallicity and density yield the same IMF independent at which redshift they are found. 
The integration over the freshly formed  
IMFs results in a gwIMF that varies with SFR and metallicity. In the IGIMF theory, gwIMF variations are driven by the physics on the embedded cluster scales. An important aspect of the IGIMF is therefore that it is automatically consistent with the stellar populations in star clusters.

The calculations presented in this work deal with the first approach and are mainly based on the publicly available python module GalIMF 
\citep{Yan2017} where the implementation of the IGIMF theory is described in more detail.  An equivalent FORTRAN package is also available (Zonoozi et al., 2018, submitted:
\href{https://github.com/ahzonoozi/GWIMF}{https://github.com/ahzonoozi/GWIMF}). 
Throughout this text we use $\log{}$ or $\log_{10}$ independently and referring always to the decimal logarithm.

\subsection{Star forming regions in a galaxy}
\label{sec:CSFEs}
Observational evidence shows that star-formation is always concentrated in small (sub-pc-scale), dense ($>10^4\,$cm$^{-3}$), and massive H2 cores within molecular clouds \citep{Tafalla2002, Wu2010}.
We refer to the star-forming cloud cores as correlated star formating events (CSFEs). Depending on their density (and thus mass), these CSFEs form from a few binaries to millions of stars. For practical purposes they can be called embedded
clusters or clumps \citep{Lada2003,Lada2010,Gieles2012,Megeath2016,Kroupa2018} even though the definition of star cluster is neither precise nor unique \citep{Bressert2010,Ascenso2018}. For example, the low- and high-density star-formation activity in the Orion~A and~B molecular clouds is organsied in such CSFEs (fig.~8 in \citealt{Megeath2016}). The important point however, independently of how these CSFEs are called, embedded clusters or just stellar groups, is that these form a co-eval (within a few $0.1\,$Myr) population of stars which can be described using the stellar IMF.  For simplicity we refer to the newly formed stellar groups/CSFEs as embedded clusters. 

\begin{figure*}[ht!] \begin{center}
\scalebox{0.6}{\includegraphics{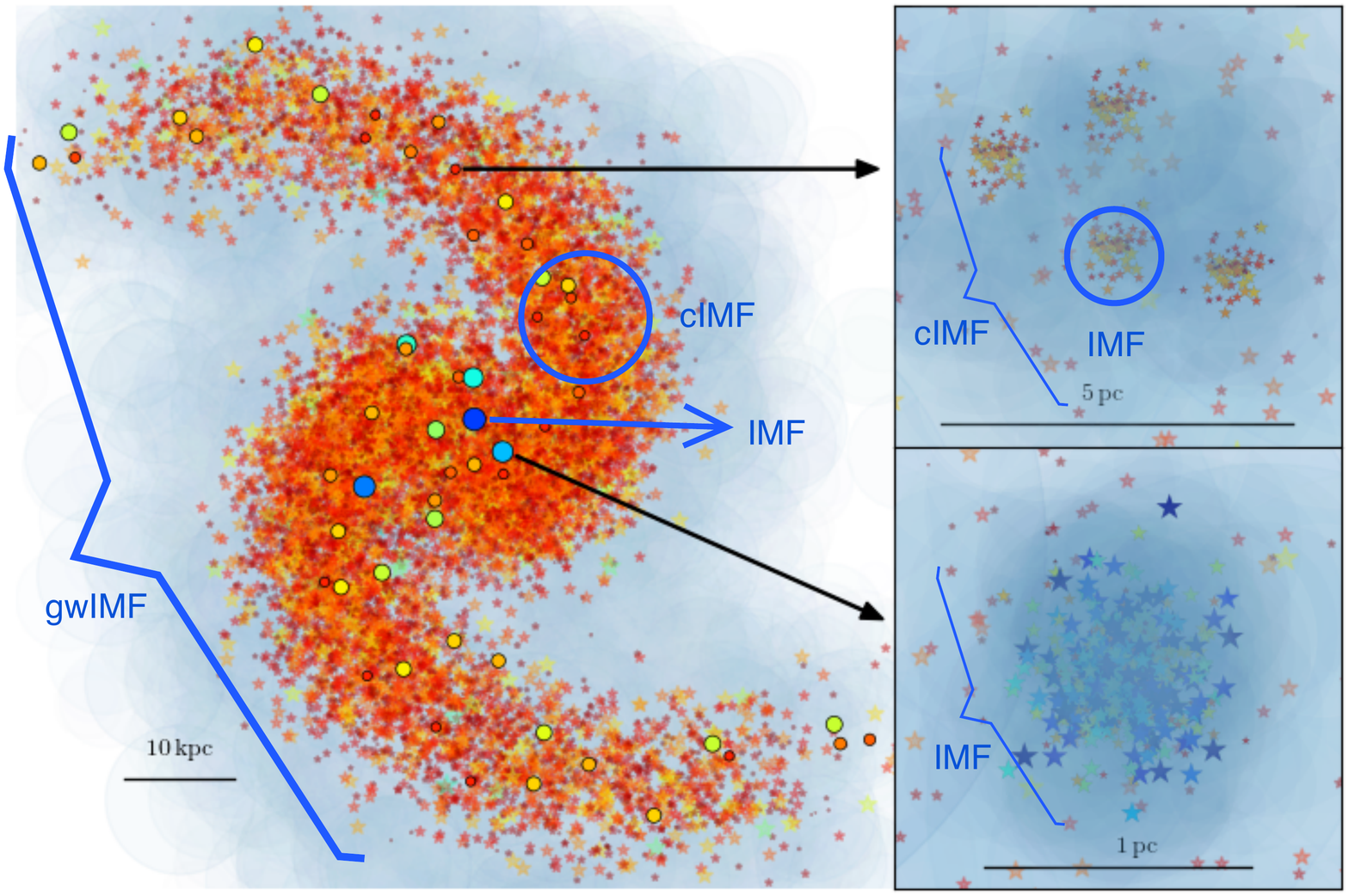}}
\end{center}
\caption{
Schematic showing a late-type galaxy. Its field population is represented by red and orange stars. The newly formed stellar population is marked by colored circles which represent 
CSFEs (embedded star clusters). The colors and sizes of symbols scale with stellar/cluster mass. The acronyms from Tab.~\ref{tab:imfs} are shown here with examples. \textbf{Right bottom panel:} A young massive embedded cluster, which will most likely survive and contribute the galaxy's open star cluster population. 
\textbf{Right top panel:} Young embedded cluster complex composed of a number of low-mass embedded clusters, which will evolve into a T-Tauri association once the embedded clusters expand after loss of their residual gas and disperse into the galaxy field stellar population.
} \label{fig:gal_sk} \end{figure*}

A visualization of a newly formed stellar population is shown as a sketch in Fig.~\ref{fig:gal_sk} 
where the right panels illustrate how different individual star forming regions can be. The massive cluster containing many O stars will most likely survive as an open cluster \citep{KAH, Brinkmann2017}. Low-mass embedded clusters or groups will on the other hand dissolve quickly due to loss of their residual gas \citep{Brinkmann2017} and energy-equipartition driven evaporation \citep{BinneyTremaine1987, HeggieHut2003, BaumgardtMakino2003}. Examples of this range of embedded clusters can be seen in Orion \citep{Megeath2016}, each having spatial dimensions
comparable to the molecular cloud filaments and the intersection
thereof \citep{Andre2016, Lu2018}. 

In general the sum of outflows and stellar radiation compensate the depth of the gravitational potential of the embedded cluster and individual proto-stars such that star-formation in the embedded clusters is feedback regulated. Indeed, observational evidence shows that the majority of gas will be expelled from massive star-forming cores (e.g., in Orion A and B the star formation efficiency is less than about 30 per cent per embedded cluster, \citealt{Megeath2016}). Observations of outlfows from embedded clusters document this in action \citep{Whitmore1999,Zhang2001, Smith2005, Qiu2007, Qiu2008,Qiu2011}. Magneto-hydrodynamical simulations \citep{Machida2012,Bate2014, Federrath2014, Federrath2015, Federrath2016} also lead to the same result. 
Well-observed CSFEs, e.g. the Orion Nebula Cluster, Pleiades, NGC3603 and R136, span a stellar mass range from a few~10 to a few $10^5\,M_\odot$ in stars. Their dynamics can be well reproduced in $N$body simulations, with star formation efficiency $\approx 33$ per cent, 10 km/s gas expulsion, and 0.6 Myr for the typical embedded phase 
\citep{Kroupa2003,KAH,Banerjee2013,Banerjee2014,
Banerjee2015,Banerjee2017}.

Observations suggest that even T~Tauri associations loose their residual gas on a time scale of about a Myr
\citep{Neuhaueser1998}, which is supported by magneto-hydro-dynamic-radiative-transfer simulations by \cite{Hansen2012}. Given the loss of about 2/3 of the binding mass,  embedded clusters expand by a factor of three to five due to the expulsion of most of their gas such that embedded clusters with a stellar mass smaller than about $10^4\,M_\odot$ loose more than 60~per cent of their stars, the rest re-virialising to form longer-lived low-mass open clusters \citep{Brinkmann2017}. This implies that embedded clusters which are typical in molecular clouds become unbound within less than a Myr, forming stellar associations if multiple embedded clusters spawn from one molecular cloud (e.g. also \citealt{Lim+18}). The observed properties of OB associations are further established by stars being efficiently ejected form their embedded clusters \citep{Oh+15,OK16}. Interesting in this context is that a recent study was able to identify a complex expansion pattern consisting of multiple expanding substructures within the OB association Scorpius-Centaurus using Gaia data \citep[][e.g. their Fig 11]{Wright2018}. 

\subsection{Assumptions}
\label{sec:assumptions}
\paragraph{1. The embedded cluster initial mass function (ECMF)\\}
The embedded cluster initial mass function (ECMF) represents the birth-star-cluster population's 
mass distribution, $\xi_{\mathrm{ecl}}$, formed in one formation time scale throughout a galaxy ($\delta t$, see Paragraph~2 below).
In the present IGIMF implementation, based on the available data (\citealt{Yan2017} and references therein),  it is assumed that the ECMF is represented by a single power law with 
a slope $\beta$ as a function of galactic SFR,
\begin{equation}
\xi_{\mathrm{ecl}}(M_{\rm ecl},\mathrm{SFR}) =    \left\{ \begin{array}{ll}
0, \hspace{1.65cm} M \le M_{\mathrm{ecl,min}} \,, \\
k_{\mathrm{ecl}} M^{-\beta(\mathrm{SFR})}, \hspace{0.1cm} M_{\mathrm{ecl,min}}\leq M_{\rm ecl} < M_{\mathrm{ecl,max}}(\mathrm{SFR}) \,, \\
0, \hspace{1.65cm} M_{\mathrm{ecl,max}}(\mathrm{SFR})  \leq M_{\rm ecl} \,, \\
\end{array} \right.
\label{eq:ECIMF}
\end{equation}
where $M_{\mathrm{ecl,min}} = 5\,M_{\odot}$ is the lower limit of the mass in stars of the embedded cluster \citep{Kirk2012, Kroupa2003}, 
$M_{\mathrm{ecl,max}}$ is the upper limit for the embedded cluster's stellar  mass, being computed within the IGIMF theory \citep[see][]{Schulz2015, Yan2017}, and $k_{\mathrm{ecl}}$ is a normalisation constant. 
If $\mathrm{d}N$ is the number of embedded cluster with masses in stars between $M_{\rm ecl}$ and $M_{\rm ecl}+\mathrm{d}M_{\rm ecl}$ values, then 
$\xi_{\mathrm{ecl}} = \mathrm{d}N_{\mathrm{ecl}}/\mathrm{d}M_{\rm ecl}$. 

The detailed shape of the ECMF might be different from the assumption 
of a single power law \citep[e.g][]{Lieberz2017}, however such a change can be easily incorporated into the IGIMF framework 
and is not expected to cause significant differences to the results presented here. 
The dependence of $\beta$ on SFR is described by the relation \citep{Weidner2004, Weidner2013b,Yan2017}, 
\begin{eqnarray}
\beta = -0.106\log_{10}{\mathrm{SFR}}+2\,.
\label{[eq:beta]}
\end{eqnarray}
This description implies that galaxies undergoing major star bursts produce top-heavy ECMFs. Observational data suggests that the ECMF may not be a probability density distribution function \citep{Pflamm2013}.

\paragraph{2. Formation time scale of the stellar population\\}
In a galaxy, in which stars are being formed over hundreds of Myr to many Gyr, it is important to 
establish the duration, $\delta t$, over which the inter-stellar medium spawns a complete populations of embedded clusters. 
This time scale 
allows us to compute the total stellar mass, $M_{\mathrm{tot}}$, formed 
within $\delta t$ as the integral over the ECMF over all embedded cluster masses, 
\begin{equation}
  \label{eq:Mtot}
  M_{\mathrm{tot}} = \mathrm{SFR} \cdot \delta t\,.
\end{equation}
Solving this integral yields $M_{\rm ecl, max}({\rm SFR})$.

We set $\delta t = 10$ Myr for several reasons:
The time scale for galaxy-wide variations of the SFR is $\approx \mathrm{few}\,100$ Myr \citep{Renaud2016}. 
$\delta t \approx 10\,$Myr corresponds to the time-scale over which molecular clouds are forming 
stars \citep{Egusa2004,Egusa2009,Fukui2010,Meidt2015} and to the survival/dissolution timescale of giant molecular clouds \citep{Leisawitz1989,Padoan2016,Padoan2017}.
In addition, it has been shown that the $\delta t\approx 10\,$Myr timescale predicts the $M_{\mathrm{ecl,max}}-\mathrm{SFR}$ relation within the 
IGIMF concept consistent with observational data \citep{Weidner2004, Schulz2015, Yan2017}.  It is to be emphasized that this time scale of $\delta t\approx10\,$Myr is neither the pre-main-sequence stellar-evolution nor the stellar-evolution time scale. Essentially, $\delta \approx 10\,$Myr is the free-fall time of bound regions of molecular clouds and the time-cycle over which the inter-stellar medium of a galaxy spawns new populations of embedded clusters. It is evident in the offsets between H$\alpha$ and CO spiral arms \citep{Egusa2004,Egusa2009}.

\paragraph{3. The stellar IMF\\}
We describe the stellar IMF as a
multi-power law function,
\begin{equation}
\xi_{\star} (m) =    \left\{ \begin{array}{ll}
k_1 m^{-\alpha_1} \hspace{1.65cm} m_{\rm min}\leq m/M_{\odot}<0.50 \,, \\
k_2 m^{-\alpha_2} \hspace{1.65cm} 0.50\leq m/M_{\odot}<1.00 \,, \\
k_2 m^{-\alpha_3} \hspace{1.65cm} 1.00\leq m/M_{\odot}< m_{\mathrm{max}} \,, \\
\end{array} \right.
\label{eq:IMF}
\end{equation}
where.
\begin{eqnarray}
    \xi_{\star}(m) = \mathrm{d} N_{\star}/\mathrm{d} m\, ,
\end{eqnarray}
is the number of stars per unit of mass and $k_i$ are normalization constants which also 
ensure continuity of the IMF,  $m_{\rm min}=0.08\,M_\odot$ is the minimum stellar mass used here and the function $m_{\rm max} = {\rm WK}(M_{\rm ecl}) \le m_{\rm max*}\approx 150\,M_\odot$ is the most massive star in the embedded cluster with stellar mass $M_{\rm ecl}$ (the $m_{\rm max}-M_{\rm ecl}$ relation, \citealt{Weidner2006}) and $m_{\rm max*}$ is the empirical physical upper mass limit of stars \citep{WeidnerKroupa2004, Figer05, OeyClarke05, Koen06,Maiz+2007}. Stars with a higher mass are most likely formed through stellar-dynamically induced mergers \citep{OK12, BKO12}.

Here we assume that star-formation is feedback self-regulated and thus we implement the $m_{\rm max}=WK(M_{\rm ecl})$ relation based on observational data \citep{Weidner2006,Kirk2012, Weidner2013b,Ramirez+16,Megeath2016,Stephens2017, Yan2017} assuming no intrinsic scatter \citep{Weidner2010,Weidner2013b}. Despite the newer data \citep[e.g.][]{Ramirez+16,Stephens2017} supporting the existence of this $m_{\rm max}-M_{\rm ecl}$ relation, future investigations of the interpretation and of the true scatter in it will be useful.

As a benchmark we use the canonical IMF $\alpha_i$ values derived from Galactic star forming
regions by \cite{Kroupa2001}, where $\alpha_1=1.3$ and $\alpha_2=\alpha_3 =2.3$ 
(the Salpeter--Massey index or slope, \citealt{Salpeter1955, Massey2003}). These are mostly based on in-depth analysis of star counts \citep{Kroupa1993} as well as young and open clusters for $m \le 1\,M_\odot$ and on the work of \cite{Massey2003} for $m>1\,M_\odot$.
The relation for $\alpha_3$, derived 
by \cite{Marks2012TH} \citep[see erratum][]{Marks2014_erratum}, is 
\begin{equation}
\alpha_3=   \left\{ \begin{array}{ll} 2.3 \hspace{2.15cm}
\mathrm{if}\,x<-0.87\,,\\ -0.41x+1.94 \,
\hspace{0.5cm}\mathrm{if}\,x\geq-0.87\,, \end{array} \right.
\label{eq:upperIMF} \end{equation} 
where \begin{equation} x =
-0.14\mathrm{[Fe/H]}+0.99\log_{10}{\left(\frac{\varrho_{cl}}{10^6M_{\odot}pc^{-3}}\right)}\,,
\label{eq:MKrel}
\end{equation}
where $\varrho_{cl}$ is the total density (gas and stars) of the embedded cluster \citep{Marks2012TH}, 
\begin{equation}
\varrho_{cl} = 3M_{cl}/4\pi r_{h}^3 \, ,
\label{eq:MR_rel}
\end{equation}
where $M_{cl}$ is initial cluster mass including gas and stars and $r_h$ is its half 
mass radius \citep{Marks2012}.  The density of the stars is expressed as, $\rho_{\rm ecl}=3\,M_{\rm ecl}/(4\pi\,r_h^3)$. We assume a star formation efficiency $33\%$ and thus 
the mass of the embedded cluster in stars, $M_{ecl}$, is $M_{ecl} = M_{cl} \cdot 0.33$. 
To estimate the value of the density, $\varrho_{ecl}$, we adopt the relation from \cite{Marks2012}, $r_h/\mathrm{pc} = 0.1M_{ecl}^{0.13}$, where $M_{ecl}$ has units of $M_{\odot}$. In addition the relation, $\log_{10}\varrho_{ecl} = 0.61 \log_{10} M_{ecl}+2.08$ allow us to formulate the relation between $\varrho_{cl}$ and $M_{ecl}$ as  $\log_{10}\varrho_{cl} = 0.61 \log_{10} M_{ecl}+2.85$.
This allows us to compute $\alpha_3$ once metallicity and the mass of star cluster is known.
From the original formulation of Eq.~\ref{eq:MKrel} by  \cite{Marks2012TH} it is possible to combine the assumptions on the cluster mass and radius to formulate the concise equation, 
\begin{equation} x =
-0.14\mathrm{[Fe/H]}+0.6\log_{10}{\left(\frac{M_{ecl}}{10^6M_{\odot}}\right)}+2.83\,.
\label{eq:MKrel2}
\end{equation}
We note that Eq.~\ref{eq:MKrel2} conveniently uses only the star cluster initial stellar mass, $M_{ecl}$, and the metallicity of the embedded cluster as input parameters.

In addition in \cite{Kroupa2002, Marks2012TH} an empirical relation for the dependence of $\alpha_{i}$, $i=1,2$, on [Fe/H] is suggested, 
\begin{equation}
  \alpha_{i} = \alpha_{ic}+\Delta \alpha \mathrm{[Fe/H]}\, ,
  \label{eq:alpha12}
\end{equation}
where $\Delta \alpha \approx 0.5$ and $\alpha_{ic}$ are the respective slopes of the canonical IMF.  This equation is based on a rough estimate by \cite{Kroupa2002} for stellar populations in the Milky Way disk, the bulge and globular clusters spanning a range of about [Fe/H]$=+0.2$ to $\approx -2$. Beyond this range the results are based on an extrapolation. This is also true for the  validity of Eq.~\ref{eq:upperIMF} which is based on Galactic-field populations and a dynamical analysis of globular clusters and ultra-compact dwarf galaxies. 

We note that we use [Fe/H] as a metallicity traces and thus these relations might be re-calibrated to use more robust full metallicity, $Z$, using self-consistent chemical evolution codes.

\begin{table}
\caption{IGIMF implementations via local variations.}              
\label{tab:imfs}      
\centering                                      
\begin{tabular}{c c c c}          
\hline\hline                        
 & $\alpha_1$ & $\alpha_2$ & $\alpha_3$ \\    
\hline                                   
\textcolor{ForestGreen}{IGIMF1} & 1.3 & 2.3 & 2.3 \\      
\textcolor{blue}{IGIMF2} & 1.3 & 2.3  & Eq. (\ref{eq:upperIMF}) \\
\textcolor{red}{IGIMF3} & Eq. (\ref{eq:alpha12}) & Eq. (\ref{eq:alpha12})  & Eq. (\ref{eq:upperIMF}) \\
\hline                                             
\end{tabular}
\tablefoot{
This table summarizes the here implemented variations of the stellar IMF resulting in different IGIMFs. The $\alpha_i$ coefficients are 
defined in Eq. (\ref{eq:IMF}). Note that the model IGIMF1 assumes the IMF to be the invariant canonical form and corresponds to the original formulation of the IGIMF theory \citep{KroupaWeidner2003,Weidner2006} before evidence for the variation  of the stellar IMF was quantified, the model IGIMF2 assumes that only the upper-end of the IMF varies with density and metallicity, while the model IGIMF3 assumes the IMF varies over all stellar masses. 
}
\end{table}

\subsection{The IGIMF formulation}
\label{sec:theIGIMF}
Based on the assumptions detailed above we can describe the stellar initial mass function for the whole galaxy, $\xi_{\mathrm{IGIMF}}$, as a sum of all the stars in all embedded clusters formed over the time $\delta t = 10\,$Myr, 
\begin{multline}
\xi_{\mathrm{IGIMF}}(m,\mathrm{SFR, [Fe/H]}) = \\ 
\int_0^{+\infty} \xi_{\star}(m,M,{\rm [Fe/H]}) 
\xi_{\mathrm{ecl}}(M,\mathrm{SFR})\mathrm{d}M\,,
\label{eq:IGIMF}
\end{multline}
where $\xi_{\mathrm{ecl}}$, the initial mass function of embedded clusters, is described by Eq.~\ref{eq:ECIMF} and the stellar initial mass function is given by Eq.~\ref{eq:IMF}, Eq.~\ref{eq:upperIMF} and Eq.~\ref{eq:alpha12}.

Eq.~\ref{eq:IGIMF} represents the general recipe for constructing gwIMF from local stellar IMFs which appear within a galaxy within the time interval $\delta t$. 
Three versions of the IGIMF are calculated (IGIMF1, IGIMF2, IGIMF3), with the properties of each being tabulated in Table~\ref{tab:imfs}.

\section{Results}

\subsection{The IGIMF grid}
Together with this publication we provide the IGIMF grid in electronic form. That is, 
for each value of the galaxy-wide SFR and [Fe/H] that is in the computed set we provide the
gwIMF (calculated as the IGIMF) in the mass range from 0.08 to 120 $M_{\odot}$.  This grid can be readily truncated at $100\,M_\odot$. The IGIMF is tabulated  as the stellar mass bin 
in one column and the 
other three columns contain IGIMF values in the form of IGIMF1/2/3 summarized in Tab.~\ref{tab:imfs}. 
The mass range is the same for the whole parameter space for easier implementation into 
any code and potential interpolation within the grid.
The IGIMF here is normalized to the total stellar mass, $M_{\mathrm{tot}}$ (Eq.~\ref{eq:Mtot}), produced in $\delta t =10\,$Myr, 
$\int_{m_{\mathrm{min}}}^{m_{\mathrm{max}}} m~\cdot~\xi_{\mathrm{IGIMF}}\,\mathrm{d}m = M_{\mathrm{tot}}$.

\begin{figure*}[ht!] \begin{center}
		\scalebox{0.8}{\includegraphics{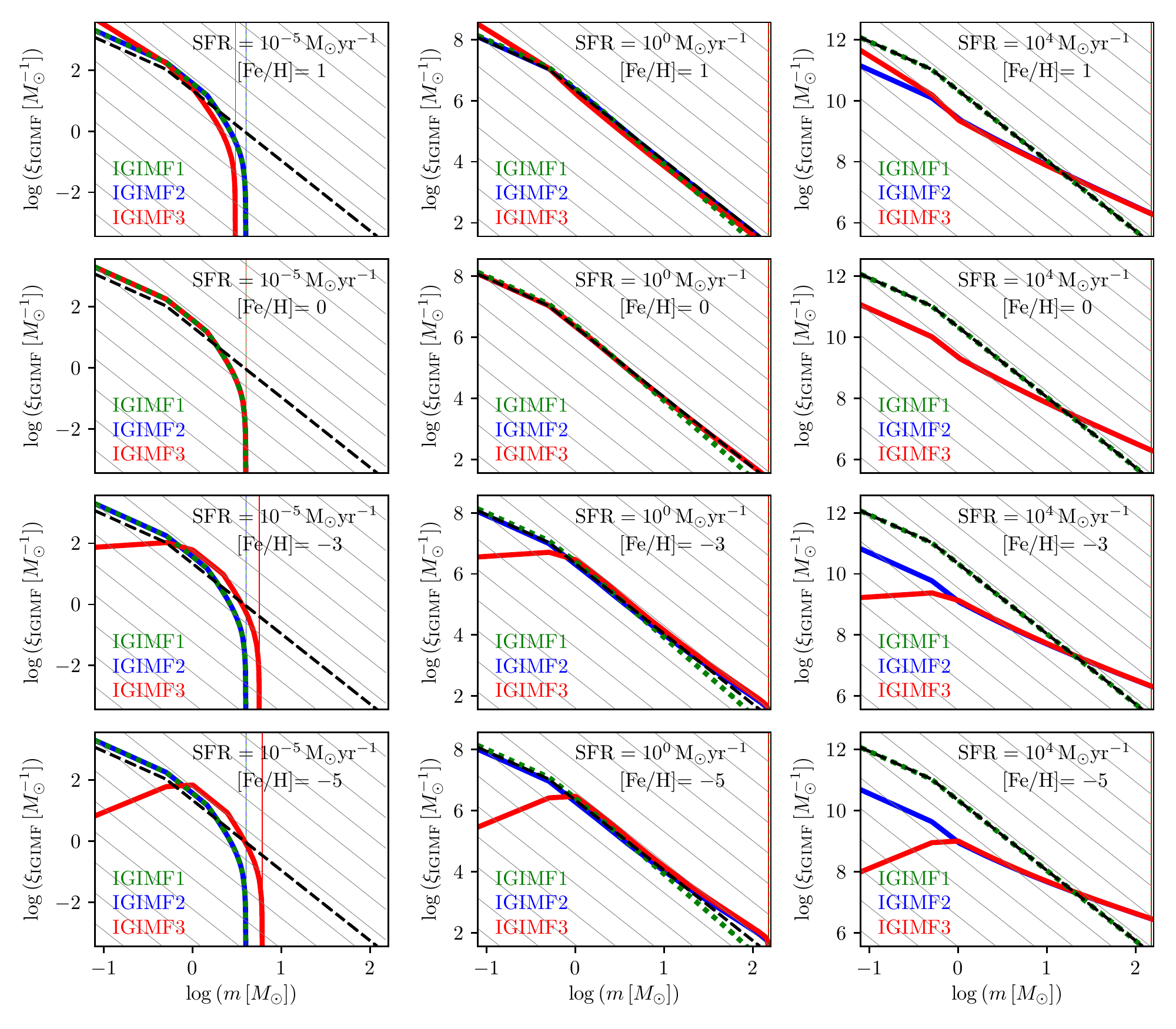}} \end{center}
	\caption{A selection of IGIMF models representing the overall characterization of the IGIMF grid published in electronic form with this work.
		The metallicity used in this figure is $\mathrm{[Fe/H]} = 1,0,-3,-5$ from top to bottom, the $\mathrm{SFR} = 10^{-5},10^0,10^4 \,\mathrm{M_{\odot}/yr}$ from left to right. 
        All IMF models are normalized to the total stellar mass formed over $\delta t = 10\,$Myr to make the comparison 
		with the canonical IMF (black dashed line in each panel) quantitative. To compare the slope variations we plot the Salpeter-Massey slope, $\alpha=2.3$, 
		as a grey-line grid in each plot.
	} \label{fig:grid} \end{figure*}

A representative selection from the grid is shown in Fig.~\ref{fig:grid} where we can see variations of the gwIMF 
over the large span of parameters. The panels on the left show that for low SFR the gwIMF is top-light. That is,
we expect a deficit of high mass stars in comparison to the canonical IMF and that the mass of the most massive star in a 
galaxy varies with metallicity due to the IMF-metallicity dependence.
For a SFR of $1\,M_{\odot}$/yr, which is approximately the SFR of the MW, the galaxy-wide IMF is very close to but slightly steeper than the canonical IMF. Therefore the IGIMF is always consistent with the MW and local star formation 
regions. It fulfills automatically this test every theory of IMF variations needs to pass. For the larger SFR values the galaxy-wide IMF
becomes top-heavy, that is relatively more massive stars form than would be given by the canonical IMF. The IGIMF1 formulation which does not implement 
IMF variations, but only the $m_{\rm max}-M_{\rm ecl}$ relation, does not show any variations at SFR $\geq 1 M_{\odot}/\mathrm{yr}$. This is essentially the IGIMF version calculated by \cite{KroupaWeidner2003}, before the constraints on IMF variations discussed above had become evident. The IGIMF3 formulation, 
which implements the full variations of the IMF with density and metallicity, can result in bottom-heavy gwIMFs at metalicities $\mathrm{[Fe/H]} > 0$, bottom-light gwIMF for $\mathrm{[Fe/H]} < 0$ independent of SFR. For SFR $< 1 M_{\odot}/\mathrm{yr}$ the gwIMF becomes top-light independent of metallicity. For SFR $> 1 M_{\odot}/\mathrm{yr}$ gwIMF becomes top-heavy, this effect becoming stronger for $\mathrm{[Fe/H]} < 0$.

All the scripts used here are uploaded to the galIMF scripts (\href{https://github.com/Azeret/galIMF}{https://github.com/Azeret/galIMF}) such that the galIMF module can be self-consistently implemented into any chemical evolution code.

A compact quantification of the changing shape of the IGIMF for different assumptions can be achieved 
by calculating the mass ratios in different stellar mass bins. To see how relevant low mass stars are to the total mass budget formed in $\delta t = 10\,$Myr, the $F_{05}$ parameter \citep{Weidner2013c} is defined as, 
\begin{equation}
F_{05} = \frac{\int_{m_{\mathrm{min}}}^{0.5\, M_{\odot}} m \cdot \xi_{\mathrm{IGIMF}}\, \mathrm{d}m}   {M_{\mathrm{tot}}}\,.
\label{eq:F05}
\end{equation}
It quantifies the fraction of stellar mass in stars less massive than $0.5\,M_\odot$ relative to the total initial stellar mass. The dependency of $F_{05}$ on the SFR and metallicity is shown in Fig.~\ref{fig:F05} for the different IGIMF formulations. Values $F_{05}>0.25$ indicate bottom-heavy IGIMFs. 
Values of $F_{05}>0.6$ are required to match IMF-sensitive spectral features in elliptical galaxies
\citep{LaBarbera+13, Ferreras+15}. Such a large $F_{05}$ values would not lead to very high dynamical mass-to-light ratios as
the resulting IGIMF is not significantly steeper than the canonical IMF for $m<0.5\,M_\odot$. This is very important, because an IGIMF with a single 
Salpeter-power-law index over all stellar masses would lead to 
unrealistically high dynamical mass-to-light ratios (see \citealt{Ferreras+13}).

Similarly, the mass-fraction of stars with $m<0.4\,M_\odot$ relative to the present-day stellar mass (in all stars less massive than $0.8\,M_\odot$) is defined as 
\begin{equation}
F_{04/08} = \frac{\int_{m_{\mathrm{min}}}^{0.4\, M_{\odot}} m \cdot \xi_{\mathrm{IGIMF}}\, \mathrm{d}m}    {\int_{m_{\mathrm{min}}}^{0.8\, M_{\odot}}    m \cdot \xi_{\mathrm{IGIMF}}\,\mathrm{d}m }\,.
\label{eq:F04p}
\end{equation}
It constitutes an approximation to a stellar population which is about 12~Gyr old. This parameter, plotted in Fig.~\ref{fig:F04p}, informs on the bottom-heaviness of the present-day stellar population ignoring stellar remnants. 
Furthermore, the parameter 
\begin{equation}
F_{8} = \frac{ \int_{8.0\, M_{\odot}}^{m_{\rm max}} m \cdot \xi_{\mathrm{IGIMF}}\, \mathrm{d}m}   {M_{\mathrm{tot}}}\,,
\label{eq:F80}
\end{equation}
is the mass fraction of stars more massive than $8.0\,M_\odot$ relative to the total initial stellar mass formed in 10~Myr, $M_{\rm tot}$, and indicates the degree of top-heaviness of the IGIMFs (Fig.~~\ref{fig:F80}).

\subsection{The evolution of the gwIMF of an elliptical galaxy and its chemical evolution}
\label{sec:gwIMFevol}
\begin{figure*}[ht!] \begin{center}
		\scalebox{0.8}{\includegraphics{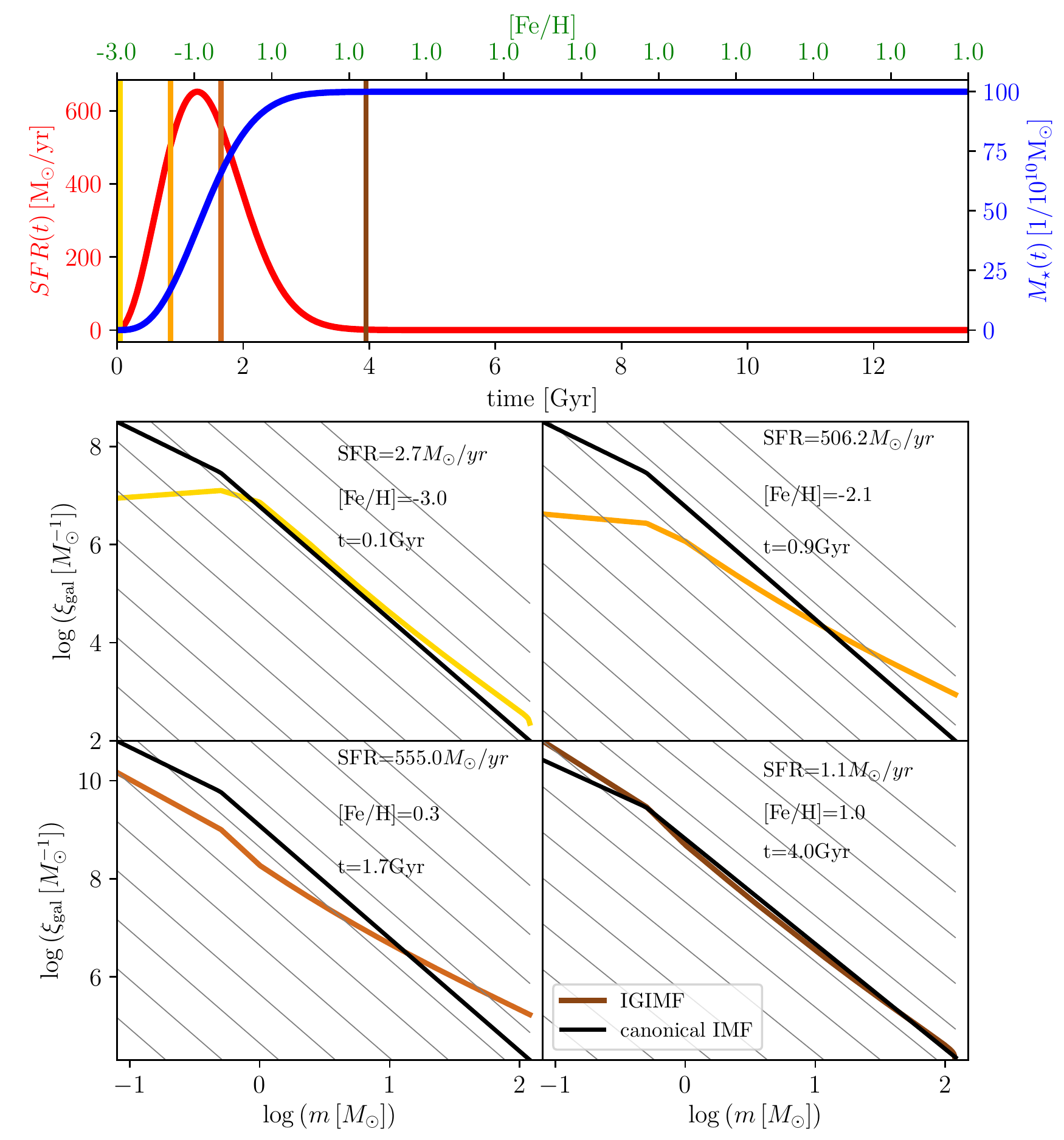}} \end{center}
	\caption{An example of IGIMF evolution with time, here plotting only the IGIMF3 model. \textbf{Top panel:} The prescribed evolution of the
		SFR with time (red solid curve, left y-axis), that is the star formation history (SFH), the stellar-mass-build-up with time (blue solid curve, right y-axis)      and also the metallicity evolution (upper x-axis). 
		This example serves to show a typical evolution and therefore the curves were synthetically created. This example of how a $10^{12}\,M_\odot$ elliptical galaxy assembles over about $1\,$Gyr is consistent with down-sizing \citep{Recchi2009}, but due to stellar evolution the stellar mass of this galaxy will be  several $\times 10^{11}\,M_\odot$ after $12\,$Gyr.
		Using the IGIMF grid, the same principle can be applied self-consistently in a chemo-dynamical code. In addition the four vertical 
		lines represent the chosen time snapshots shown in the bottom panels. \textbf{Bottom panels:} four IGIMF plots at the chosen times 
		(see top panel), showing how the IGIMF can potentially evolve throughout elliptical galaxy assembly. 
		Shown is the top-heavy phase, but also the bottom-heavy one during the metal-rich part of the evolution.
	} \label{fig:tz} \end{figure*}

\begin{figure}[ht!] \begin{center}
		\scalebox{1.0}{\includegraphics{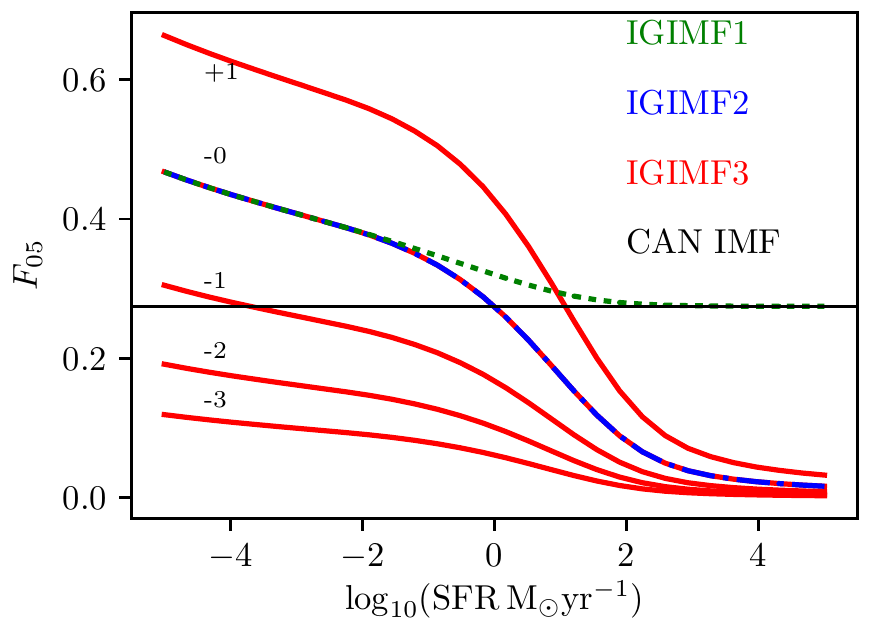}} \end{center}
	\caption{The mass-fraction formed in $\delta t=10\,$Myr in stars less massive than $0.5\,M_\odot$ relative to the total initial mass in the IGIMF models (Eq.~\ref{eq:F05}) is shown in dependence of the SFR and [Fe/H]. The canonical IMF is depicted as the horizontal black line ($F_{05}\approx 0.25$). Note that the IGIMF1 and IGIMF2 models lead to F05 values as if the gwIMF were bottom-heavy in comparison with the canonical IMF. However this is due to the normalization caused by the IGIMF being top-light for low SFRs ($< 1 M_{\odot}/\mathrm{yr}$).
	} \label{fig:F05} \end{figure}

\begin{figure}[ht!] \begin{center}
		\scalebox{1.0}{\includegraphics{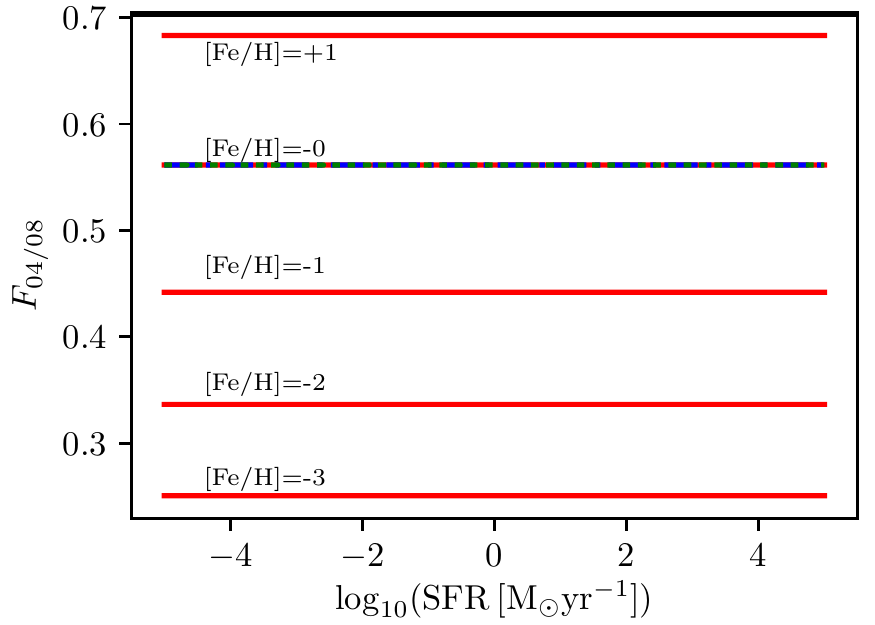}} \end{center}
	\caption{The mass-fraction formed in $\delta t=10\,$Myr in stars less massive than $0.4\,M_\odot$ relative to the total present-day stellar mass formed in $\delta t=10\,$Myr in the IGIMF models (Eq.~\ref{eq:F04p}) is shown in dependence of the SFR and [Fe/H]. The canonical IMF is depicted as the horizontal black line ($F_{04/08}\approx 0.7$), as are the IGIMF1 and IGIMF2 models as these are metallicity independent. The present-day stellar population is assumed to contain only stars less massive than $0.8\,M_\odot$, ignoring remnant masses. The fractions are constant because these IGIMF models do not depend on the SFR for stars with $m<1\,M_\odot$.
	} \label{fig:F04p} \end{figure}

\begin{figure}[ht!] \begin{center}
		\scalebox{1.0}{\includegraphics{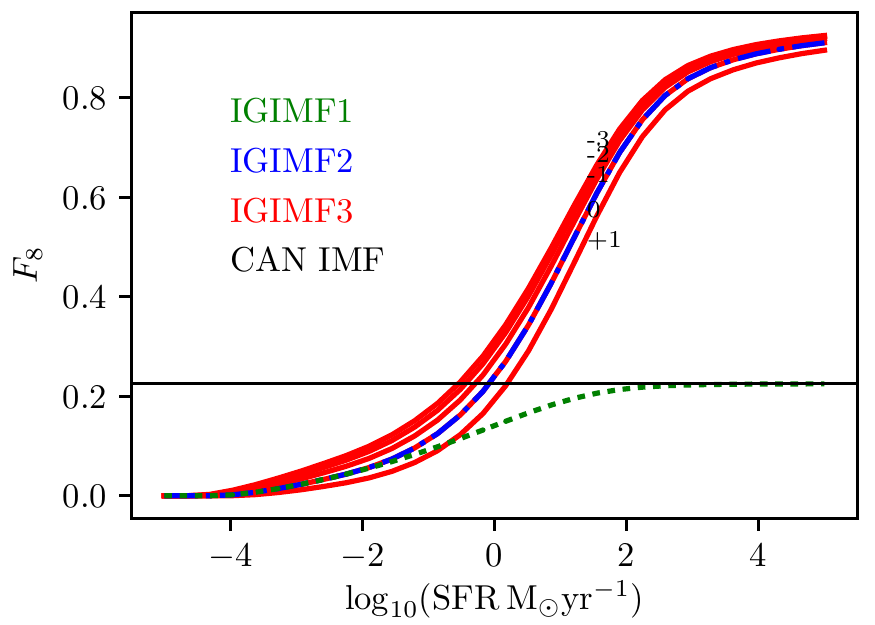}} \end{center}
\caption{The mass-fraction formed in $\delta t=10\,$Myr in stars more massive than $8.0\,M_\odot$ relative to the total initial mass in the IGIMF models (Eq.~\ref{eq:F80}) is shown in dependence of the SFR and [Fe/H]. The canonical IMF is depicted as the horizontal black line ($F_{80}\approx 0.22$).
	} \label{fig:F80} \end{figure}

The presented IGIMF grid, or the script using galIMF to produce the grid, can be readily implemented into  
galaxy chemical evolutionary codes to obtain a self-consistent galaxy-wide IMF evolution with time. 
To show that the IGIMF approach is promising in this regard, we created a burst star-formation history that approximately resembles the formation of an elliptical galaxy with a total mass in all stars formed of $10^{12}\,M_\odot$. Its present-day, about 12~Gyr old counterpart, would, according to the present results (Fig.~\ref{fig:F05}), have a mass of about $2\times 10^{11}\,M_\odot$ in main-sequence stars.
The [Fe/H] enrichment, a prescribed function of time used here solely for the purpose of demonstration,  is shown in the 
top panel in Fig. \ref{fig:tz}. For each $\delta t=10\,$Myr epoch the IGIMF is computed for the given SFR and 
[Fe/H] value. The bottom set of panels show the evolution of the IGIMF.
In this example, the gwIMF is 
top-heavy at high SFR and it
becomes bottom-heavy during the metal-rich phase of the evolution. That is, the stellar population as described by this IGIMF, can produce rapid $\alpha$ element enrichment in a 
fast first phase and can potentially produce also an overabundance of low mass stars mainly in the most metal rich center, since it is plausible that star formation may continue near the center in the high-density metal-enriched gas which has the shortest cooling time there. 

The IGIMF grid is now ready to be implemented into various chemo-dynamical codes to be tested against data in a self-consistent way. 

\subsection{Correction to the SFR--H$\alpha$ relation}
\label{sec:correction}
Given the gwIMF varies with the SFR and metallicity of a galaxy, it is expected that any observational tracer of this SFR will need to take this into account. In the following we distinguish between the true physical SFR of a galaxy, SFR, i.e. the actual mass per unit time which is being converted to stars, versus the observationally derived SFR (e.g. SFR$_{\rm K}$ in Eq.~\ref{eq:Kenn} below) which requires a tracer such as the H$\alpha$ flux which is an often used measure of the SFR subject to an assumption concerning the shape of the gwIMF. This measure works in principle by counting the number of photons emitted from recombining hydrogen atoms such that each recombination accounts for an ionising event so that the H$\alpha$ flux is a measure of the flux of ionising photons. By measuring the H$\alpha$ flux a measure of the number of massive stars which have formed is thus obtained. By assuming an IMF the total amount of mass converted to stars can be calculated.

A widely used relationship between the galaxy-wide SFR and the measured integrated H$\alpha$ flux of a galaxy is given by \citet[][his Eq. (2)]{Kennicutt1998}, 
\begin{equation}
	\mathrm{SFR}_{\rm K}/(M_{\odot}\mathrm{yr}^{-1}) = 7.9\cdot 10^{-42} L(\mathrm{H}\alpha) / (\mathrm{ergs\cdot s^{-1}})\, ,
    \label{eq:Kenn}
\end{equation}
being derived based on a single power-law IMF having a Salpeter slope \citep{Salpeter1955} in the mass range (0.1, 100) $M_{\odot}$ 
assuming Solar metallicity. 
Therefore this relation needs to be corrected for any IMF variation and metallicity, as has been done already previously \citep[e.g.][]{Lee2002,Lee2009,Pflamm2007}.

We note the following:
For low SFR$\,\lessapprox 1\,M_{\odot}/$yr and under the assumption of the universal IMF being a probability density distribution function, the Kennicutt SFR$_{\rm K}$--H$\alpha$ relation (Eq.~\ref{eq:Kenn}), when applied to star-forming systems constrained by an ECMF with $M_{min} < m_{max}$, large fluctuations in the measured SFRs would be obtained. The average SFR would also be biased to smaller values because the galaxy would typically lack massive stars. Such a bias can be larger than 0.5 dex for $\log_{10}\left({SFR/(M_{\odot}/\mathrm{yr})}\right) \lessapprox -4$, assuming $M_{min} = 20\,M_{\odot}$ due to the combination of stochastic effects and the ECMF constraint 
\citep[e.g.][]{daSilva2014}.
\footnote{\label{foot:SLUG}
The stochastically lighting up galaxies (SLUG) approach \citep{daSilva2014} is an alternative (to the newer IGIMF formulation) which implements star formation basically using principles from a first formulation of the IGIMF theory \citep{KroupaWeidner2003,Weidner2005,Weidner2006}. In SLUG, a galaxy is constructed by first drawing the masses of star clusters. These star clusters are then filled up with stars by randomly choosing from an IMF until the cluster mass is reached \citep[][their Sec. 2.1]{daSilva2014}. Therefore this is not fully random sampling but mass-constrained sampling instead \citep{Weidner2006}. This means that it is impossible to have a $100\,M_\odot$ star in a $80\,M_\odot$ star cluster, for example. 
The SLUG approach therefore is along the original IGIMF-approach (which fundamentally rests on the notion that the stellar population in a galaxy is formed in star clusters and thus constitutes conditional stochasticity, \citealt{KroupaWeidner2003,Weidner2006}). We note however that the SLUG approach differs from the current IGIMF formulations by not explicitly accounting for the $m_{\max}-M_{\rm ecl}$ relation (since each cluster is assumed to be populated randomly by stars subject to the mass constraint) while the published IGIMF work fulfills this constraint. 
Whether there is a physical function $m_{\rm max}=WK(M_{\rm ecl})$ is thus an important problem to continue to study. 
The SLUG approach therefore does not comprise pure stochastic sampling from a galaxy-wide IMF which would yield no deficit in ionising stars on average for galaxies with low SFRs \citep{Pflamm2009}. Instead, in SLUG (as in a original form of the IGIMF theory), a deficit comes about because of the mass-constraint on the stars within a cluster imposed by the masses of the star clusters. SLUG is thus a useful tool for comparison with the IGIMF theory as applied here which is deterministic, being related to the concept that star formation is  strongly feedback-self-regulated, by imposing the additional condition (in addition to the cluster mass $M_{\rm ecl}$) that the $m_{\rm max}-M_{\rm ecl}$ relation be obeyed.  We also note that similar ideas to those underlying the IGIMF theory have been considered by \cite{vanBeveren1982}.}

Here, corrections of the  SFR--H$\alpha$ function are presented in the full IGIMF (IGIMF1 and IGIMF3) framework for the first time. Note that the IGIMF2 models yield the same results as the IGMF3 models for Solar metallicty and is metallicity independent.
For this purpose the galIMF module is linked with the PEGASE stellar population synthesis code
(\citealt{Fioc2011}, see also \citealt{Fioc1999} for an astro-ph documented manual)
taking advantage of the PyPegase python wrapper\footnote{https://github.com/coljac/pypegase}. 
The H$\alpha$ flux is computed by PEGASE directly from the ionizing photons. Even though it is possible to introduce, for example, dust as an absorber, we do not use any additional parameters in our computations.
The PEGASE code is structured such that it does not allow the input IMF or gwIMF to vary during the computation. In our application the gwIMF however varies with the SFR.  Thus we limited our simulations to ones with a constant SFR and metallicity over the time-scale $\delta t = 10\,$Myr. The gwIMF computed by the IGIMF theory is a continuous function which is approximated by multi-slope power-law functions which are translated to an input file for PEGASE. In practice, we use four slopes to describe the calculated IGIMF: a power-law fit to each of the four mass ranges $0.08-0.5\,M_\odot$, $0.5-1.0\,M_\odot$, $1- 0.8m_{max}\,M_\odot$,
$0.8m_{max}-150\,M_{\odot}$. This four-segment power-law description provides an excellent approximation to the full IGIMF over all stellar masses. Nevertheless, it would be better if the full numerical form of the IGIMF can be used for such calculations with PEGASE in the future (PEGASE does not currently enable an IMF to be read-in as a data file but requires the IMF to be defined as power-law sections).

This allows us to calculate the H$\alpha$ flux as an output from the PEGASE code for a chosen gwIMF and metallicity and thus to quantify the SFR--H$\alpha$ relations for the metallicity dependent formulation of the IGIMF. From this the correction for each metallicty with respect to the Kennicut SFR$_{\rm K}$--H$\alpha$ relation can be computed.

\begin{figure*}[ht]
	\scalebox{1.0}{\includegraphics{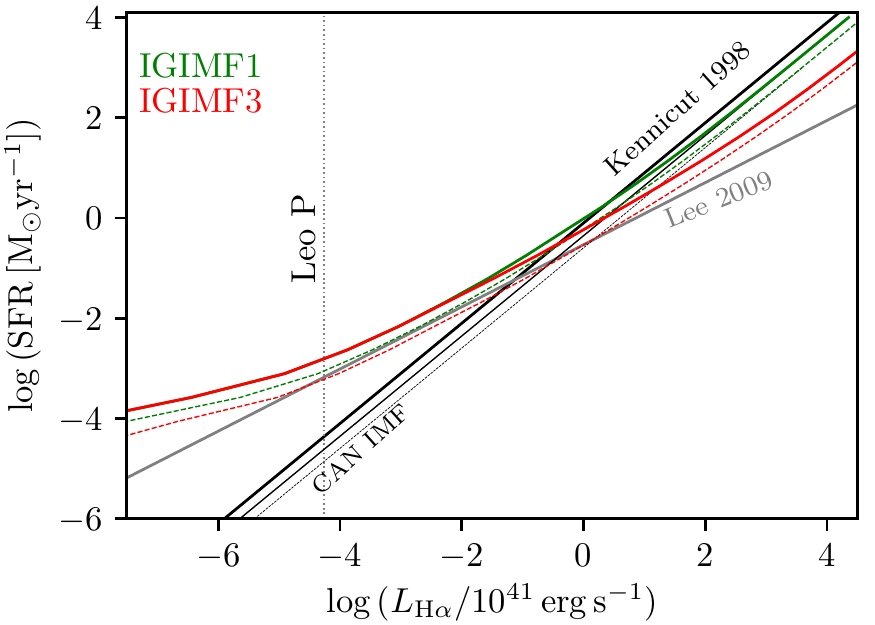}}
	\scalebox{1.0}{\includegraphics{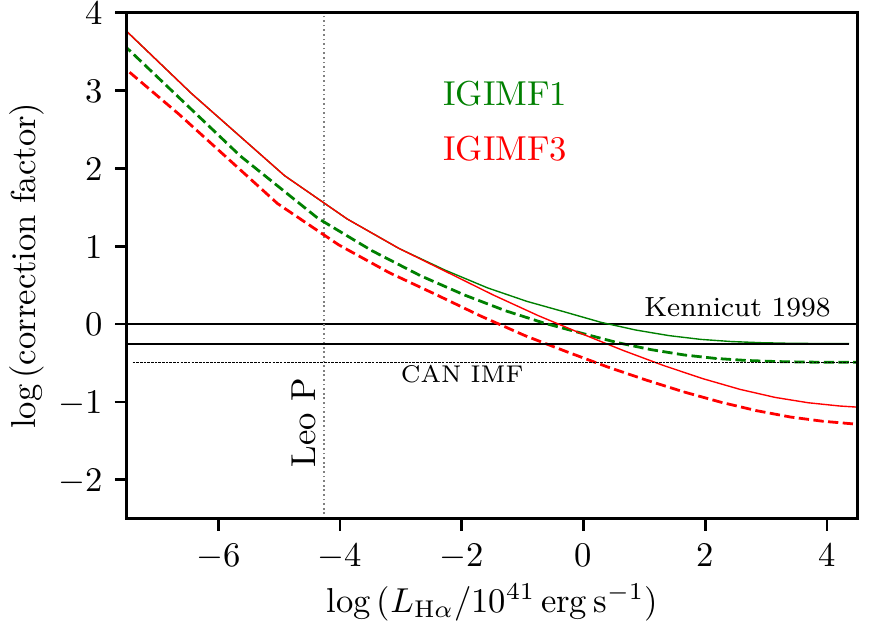}}
	\caption{
	\textbf{Left panel:}The SFR--H$\alpha$ relation. The IGIMF version in comparison with the empirical power laws proposed by \cite{Kennicutt1998} and \cite{Lee2009}. The red and green solid lines are IGIMF1/3 computed for solar metallicity [Fe/H]$=0$ and the dashed lines represent the case of sub-solar metallicity, [Fe/H]$=-2$.
    \textbf{Right panel:} Corrections to the Kennicutt SFR$_{\rm K}$--H$\alpha$ relation (Eq.~\ref{eq:corr}). The true (IGIMF) SFRs are divided by the Kennicutt value. 
    The H$\alpha$ flux of LeoP is shown by the vertical dotted line in both panels. For example, from the left panel it is evident that the IGIMF1,3 models yield values of the SFR that are consistent with the presence of massive stars in the Leo P galaxy. The two panels show that dwarf galaxies wth H$\alpha$ fluxes near $10^{36}\,$erg/s have a SFR which is more than 100 times larger than that suggested by the Kennicutt relation (Eq.~\ref{eq:Kenn}), while massive or profusely star-forming galaxies with H$\alpha\approx 10^{44}\,$erg/s have SFRs which can be 10--100~times smaller than given by the traditional Kennicutt relation. 
As for the left panel the solid lines are for solar metallicity [Fe/H]$=0$ and the dashed lines are for [Fe/H]$=-2$.
			} \label{fig:sfr} \end{figure*}

The Solar-metallicity SFR--H$\alpha$ relations are shown in the left panel of Fig.~\ref{fig:sfr}, and the sub-Solar metallicity case is shown in 
the right panel. In addition to the Kennicutt SFR$_{\rm K}$--H$\alpha$ relation and the IGIMF1,3 ones, we show the empirical correction of this relation proposed by \cite{Lee2009} based on far ultraviolet (FUV) non-ionizing continuum and H$\alpha$ nebular emission, which deviates from the Kennicutt SFR$_{\rm K}$--H$\alpha$ relation and is closer to the IGIMF relation. 

For the purpose of general use of the corrected relations in~Fig.~\ref{fig:sfr}, 
the IGIMF SFR--H$\alpha$ relations are represented with $3^{\mathrm{rd}}$ order polynomials,
\begin{eqnarray}
\log_{10}(\mathrm{SFR}_{\rm IGIMF,i}/(M_{\odot}\mathrm{yr}^{-1})) = a x^3 + b x^2 +c x^1 + d  \, ,
\label{eq:SFR_Halpha}
\end{eqnarray}
where $i=1,3$ and $x=\log_{10}(L_{\mathrm{H\alpha}} / (\mathrm{ergs\cdot s^{-1}}))$. The polynomial coefficients for different metallicities and for the IGIMF1,3 models are summarized in Tab.~\ref{tab:fit}. The sub-Solar values are consistent with the results of \cite{Boquien2014}.

The correction factor (Fig.~\ref{fig:sfr}) is calculated as follows:
\begin{equation}
{\rm corrections\;factor}({\rm H}\alpha) = \frac{\mathrm{SFR}_{\rm IGIMF,i}(L_{\mathrm{H\alpha}})}
{\mathrm{SFR}_{\rm K}(L_{\mathrm{H\alpha}})}.
\label{eq:corr}
\end{equation}

\begin{table*}
\caption{Coefficients of the fits of eq.~(\ref{eq:SFR_Halpha}) to the SFR--H$\alpha$ relations and to the 
correction factor, see Fig.~\ref{fig:sfr}.}              
\label{tab:fit} 
\centering 
\small
\begin{tabular}{c c c c c c} 
\hline\hline 
SFR--H$\alpha-$ relation & & & & & \\ 
\hline
IGIMF & [Fe/H] & $a$ & $b$ & $c$ & $d$\\   
\hline
\textcolor{ForestGreen}{IGIMF1} & 0 & $7.123 \cdot 10^{1}$ & $-5.45819 \cdot 10^{0}$ & $1.1949 \cdot 10^{-1}$
& $-7.0 \cdot 10^{-4}$\\
\textcolor{ForestGreen}{IGIMF1} & -2 & $\,\,1.038596 \cdot 10^{2}$ & $-7.98057 \cdot 10^{0}$ &  
$1.838 \cdot 10^{-1}$ &$1.25 \cdot 10^{-3}$ \\
\textcolor{red}{IGIMF3} & 0 & $2.8893 \cdot 10^{0}$ & $-4.6408 \cdot 10^{-1}$ & 
$4.1 \cdot 10^{-4}$ & $2.2 \cdot 10^{-4}$\\
\textcolor{red}{IGIMF3} & -2 & $2.394696 \cdot 10^{1}$ & $-2.2008 \cdot 10^{0}$ &  
$4.623 \cdot 10^{-2}$ & $-1.7 \cdot 10^{-4}$\\
 \hline
 correction factor  & & & & \\ 
 \hline
 \textcolor{ForestGreen}{IGIMF1} & 0 & $4.724408 \cdot 10^{1}$ & $-1.59889 \cdot 10^{0}$ & 
 $-1.15 \cdot 10^{-3}$& $-2.91 \cdot 10^{-4}$ \\
\textcolor{ForestGreen}{IGIMF1} & -2 & $3.200615 \cdot 10^{1}$ & $-3.0079 \cdot 10^{-1}$ & 
$-3.690 \cdot 10^{-2}$ & $6.15 \cdot 10^{-4}$\\
\textcolor{red}{IGIMF3} & 0 & $6.504733\cdot 10^{1}$ & $-3.2008 \cdot 10^{0}$ & 
$4.623 \cdot 10^{-2}$ & $-1.74 \cdot 10^{-4}$\\
\textcolor{red}{IGIMF3} & 2 & $4.398965 \cdot 10^{1}$ & $-1.46407 \cdot 10^{0}$ & 
$4.1 \cdot 10^{-4}$ & $2.2 \cdot 10^{-4}$\\
\end{tabular}
\tablefoot{The fits and the precision of the coefficients gives values of the $\log$ of the SFR with precision 
of approximately 0.1 dex. }
\end{table*}

\subsubsection{The case of the Leo P galaxy}
Leo P is a late-type dwarf galaxy approximately  at a distance of 1.6 Mpc which has a metallicity [Fe/H]$\approx -1.8$ and an H$\alpha$ flux, $L_{H\alpha} = 5.5 \cdot 10^{36} \,\mathrm{ergs\cdot s^{-1}}$.  This flux comes from one HII region powered by one or two stars with individual masses of $m \approx 25\,M_{\odot}$ \citep[e.g.][]{McQuinn2015}.  

We use the measured H$\alpha$ flux as a star formation indicator with the newly-developed SFR indicators of Sec.~\ref{sec:correction}. Table~\ref{tab:leoP} summarizes the computed SFRs based on different assumptions. The masses of the most massive star and of the second-most massive star are calculated for the IGIMF1 and IGIMF3 models (IGIMF2 is indistinguishable from IGIMF3, see also \citealt{Yan2017}) using the values of the SFR derived from the observed H$\alpha$ luminosity. The most-massive star has a mass of $23-26\,M_{\odot}$, the second-most massive star has a mass in the range $16-20 \,M_{\odot}$. That is, according to the IGIMF theory, the SFR$_{\rm IGIMF, i}$ of Leo~P would be significantly larger than the standard value, SFR$_{\rm K}(L_{{\rm H}\alpha})$, being consistent with the presence of the observed massive stars in Leo~P. The SLUG approach (see footnote~\ref{foot:SLUG}) also implies a larger true SFR than given by the standard value (fig.~3 in \citealt{daSilva2014}).

The message to be taken away from this discussion is that when the H$\alpha$ flux is used  as a star-formation indicator in order to test the IGIMF theory also the appropriate H$\alpha$ SFR relation needs to be employed. That there is a physical limit to the SFR if a galaxy forms a single star only at a given time had already been emphasized by (\citealt{Pflamm2007}, their Eq.~16) who also point out that distant late-type dwarf galaxies are likely to have H$\alpha$-dark star formation. In the limit where only few ionising stars form, the UV-flux derived SFRs are more robust and these are indeed consistent with the higher SFRs as calculated using the IGIMF1 formulation as shown explicitly in fig.~8 of \cite{Lee2009}, who compare UV and H$\alpha$ based SFR indicators for dwarf galaxies for which the original IGIMF formulations (IGIMF1, which did not include the IMF variation of \citealt{Marks2012TH}) remain valid. 
We add a note of caution that part of the discrepancies between the H$\alpha$ and UV based SFR indicators may be influenced by several physical effects, such as the different gas phases (such as the diffuse inoized gas present in galaxies), photon leakage form HII regions, gas and dust abundance. We refer the reader to \cite{Calzeti2013} for a more detailed discussion of various SFR tracers and their interrelations.

\cite{McQuinn2015} constructed 
the optical color-magnitude diagram (CMD) for Leo~P in order to infer its star formation history (SFH) and assumed the invariant canonical IMF for this purpose, i.e. the authors assumed the caninvgwIMF hypothesis of Table~\ref{tab:acr}. An issue worthy of future study is to quantify the degeneracies between the shape of the gwIMF and the derived SFH. Unfortunately, the implementation of a variable gwIMF into the time-dependent scheme that would allow the self-consistent modeling of the SFH while reproducing the full CMD has not been done for the IGIMF theory yet. Knowing the SFR$_{\rm IGIMF,i}$ within the IGIMF theory using the H$\alpha$ flux allows us to discuss possible effects in the CMD and the whole low-mass stellar population in the galaxy, as is touched upon in the next section.

\begin{table}
\caption{The SFR of Leo P based on the observed H$\alpha$ flux. Note the more than one order of magnitude difference in the calculated SFR between the IGIMF models and the invariant IMF models. The \cite{Lee2009} value is derived assuming their purely empirical correction to the Kenniutt relation (reached without an underlying model for an IMF variation) which is consistent with the IGIMF models in the regime $-4 < {\rm log}_{10}(L_{H\alpha}/(10^{41}{\rm erg\,sec}^{-1}) ) < 0$. To form a single massive star with a main sequence mass of $25\,M_\odot$ over $10^5\,$yr 
\citep{Wuchterl2003,Duarte13}, a SFR$\,= 25/10^5\,M_\odot$/yr$\;=10^{-3.60}\,M_\odot$/yr is needed over this time, but a dwarf galaxy would be forming other stars as well, such that the SFRs calculated using the IGIMF theory are consistent with the existence of such a star in Leo P.}              
\label{tab:leoP} 
\centering 
\small
\begin{tabular}{l c } 
\hline\hline 
SFR--H$\alpha$ relation & $\log_{10}$({\rm SFR}/($M_{\odot}$/yr))\\
\cite{Kennicutt1998} & -4.4 \\
\citet{Lee2009} & -3.2\\
CAN IMF, [Fe/H]=0 & -4.6 \\
CAN IMF, [Fe/H]=-2 & -4.9\\
IGIMF1, [Fe/H]=0 & -2.8 \\
IGIMF1, [Fe/H]=-2 & -3.0\\
IGIMF3, [Fe/H]=0 & -2.8\\
IGIMF3, [Fe/H]=-2 & -3.2\\
\hline
\end{tabular}
\end{table}

\subsubsection{Implications for the BTFR and the CMD:}
Given the higher SFRs in the IGIMF theory produced by the top-light gwIMF, the position of 
Leo P and of other dwarf galaxies in the baryonic Tully-Fisher relation \citep[BTFR,][]{McGaugh00, Lelli+16} needs be considered 
as a consistency check. That is, if there is substantial dark star formation it might alter the total mass of the galaxy,  assuming an age. 

This problem is relevant also for the dark matter problem and notably for Milgromian gravitation (MOND, \citealt{Milgrom1983, FM12}). The application of the IGIMF theory to dwarf galaxies has already shown \citep{Pflamm2009b} that the build-up times of the observed stellar populations (as assessed using the luminosity) is well accounted for within less than a Hubble time (see their figs.10 and 11), solving the problem according to which such galaxies need longer than a Hubble time to form their stellar content if the SFR was not significantly larger in the past. Applying the IGIMF theory to dwarf galaxies therefore does not change their baryonic masses, it merely shortens their gas-consumption time-scale \citep{Pflamm2009b} and allows them to form their stellar populations within a Hubble time. The BTFR therefore remains untouched.

For the case of Leo P,
the known extent and baryonic matter in Leo P and the flat (non-rising) part of the rotational curve are prone to uncertainty  and therefore more observational data are required \citep{Giovanelli2013} to constrain the position of Leo P in the BTFR.

Another consistency test is to study if the observed colour-magnitude diagram (CMD) of Leo P can be reproduced within the IGIMF theory. This needs further work and it is to be noted that the central stellar population is similar to a canonical one (e.g. in the IGIMF theory the central embedded cluster which formed the two~$25\,M_\odot$ stars is, by construction, canonical for Solar metallicity). A detailed calculation and comparison with the observed CMD needs to resort to the local IGIMF formulation (Sec.~\ref{sec:localIGIMF}) which allows the spacial integration of stellar populations within a galaxy \citep{PflammKroupa2008}.

\subsection{Ultra-faint dwarf galaxies}
\label{sec:UFDs}
 Recent measurements by \cite{Gennaro2018} of ultra-faint dwarf (UFD) satellite galaxies with the HST, as an extension of the study by \cite{Geha2013}, suggest a possible
gwIMF variation in these galaxies in the stellar-mass range $(0.4-0.8\, M_{\odot})$. 
The authors, however, mentioned that a larger data sample is needed to improve the reliability of the presented results. In this work we use the gwIMF variations derived by \cite{Gennaro2018} as an illustrative case to show how the gwIMF variations can constrain the stellar IMF on star cluster scales using the IGIMF approach. But in order to draw more robust conclusions and firmer constraints on the low-mass end of the IMF slopes ($\alpha_1, \alpha_2$) further measurements in such objects are required.

\begin{figure}[ht]
				\scalebox{1.0}{\includegraphics{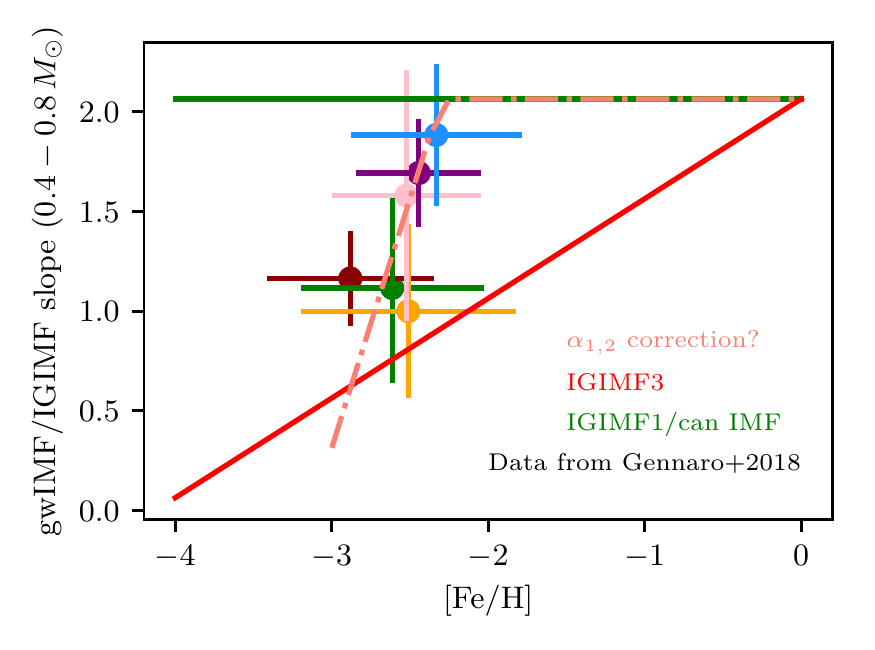}}
			\caption{
            A variable gwIMF in the stellar mass regime ($0.4-0.8\,M_{\odot}$) for~6 UFDs shown by color points from \cite{Gennaro2018}. IGIMF1 models (given by two slopes $\alpha_1 = 1.3$ and $\alpha_2 = 2.3$) are represented by the horizontal green line. These have the same effective invariant slope as the
canonical IMF in this mass regime. In red is the IGIMF3 metallicity-dependent effective slope based on Eq.~\ref{eq:alpha12}. These data may indicate the necessity for a different dependency of the $\alpha_{1,2}$ indices on metallicity than represented by Eq.~\ref{eq:alpha12}. This different dependency is indicated by the orange dot-dashed line (Eq.~\ref{eq:alpha_cor}). 
} \label{fig:local_cor} \end{figure}

Fig.~\ref{fig:local_cor} shows  inferred/measured values by \cite{Gennaro2018} in comparison with the canonical IMF and the IGIMF formulations as defined in Sec.~\ref{sec:theIGIMF}. 
\cite{Gennaro2018} assumed that the gwIMF can be reprsented by a single power-law form to derive the slope of the gwIMF in the mass range $0.4-0.8\,M_{\odot}$. To be able to compare these measurements with the 2-part power-law in the IGIMF parametrization, we compute the single power-law fit to the IGIMF/canonical IMF in the same mass range.
We can see that the IGIMF predictions do not describe the \cite{Gennaro2018} data well although the general trend is reproduced. The local IMF variations used here to calculate the IGIMF models are based on an extrapolation from data values in the range [Fe/H] $\in$ (-0.5,0) \citep{Kroupa2001,Marks2012TH}. Based on this new observation of the gwIMF in the low-mass regime  at metallicities in the range [Fe/H] $\in$ (-3,-2), a refinement of Eq.~(\ref{eq:alpha12}) may be needed, 
\begin{equation}
\alpha_{1,2}^{\mathrm{cor}} = \alpha_{1\mathrm{c},2\mathrm{c}} + \Delta \alpha^{\mathrm{cor}} \left(\mathrm{[Fe/H]+2.3}\right)\,,
\label{eq:alpha_cor}
\end{equation}
where $\Delta \alpha^{\mathrm{cor}}\approx 2.5$. 
For [Fe/H]$>-2.3$ the canonical IMF would be valid in this formulation. 
By having additional data covering a larger metallicity range and testing the robustness of these results given the uncertainties, it may be possible to identify a systematic variation of the local IMF for low-mass stars. Any such new constraints must, however, be consistent with the observationally derived stellar mass functions in present-day GCs. 

As a caveat we note that additional factors affect the empirically determined present-day index in the UFDs. For example, the fraction of unresolved multiple systems may be different in the dwarfs as it depends on the dynamical history of the population \citep{Marks2012,Marks2012TH}. Also, the formation of the stellar population in embedded clusters which expel their residual gas leading to expanding low-mass stellar populations which may be lost from a weak UFD potential may affect the finally deduced index. This process is exaggerated if the embedded clusters formed mass segregated \citep{Haghi2015}.
In addition to this the gwIMF slopes of \cite{Gennaro2018} are sensitive to the mass of the lowest stellar mass that is measured and to the form of the gwIMF that is assumed. In their next study, \cite{Gennaro2018b} use a two-part power-law gwIMF for the case of the Coma Berenices UFD finding a smaller variation with respect to the Milky Way.

\section{Discussion}

\subsection{The local or regional cIMF}
\label{sec:localIGIMF}
The gwIMF variation is parametrized within the IGIMF framework with global galaxy properties, namely the total SFR and the average metallicity. As an output we obtain the total stellar population formed in $10\,\mathrm{Myr}$ without any information about its spatial distribution in a galaxy. In reality, however, the gas density and metallicity varies spatially and therefore a mathematical formulation of the composite IMF  
(cIMF, Tab.~\ref{tab:tab1}), which takes into account the local gas surface density and metallicity at some position within the galaxy, is needed. A cIGIMF version has been formulated by \cite[][the "local" IGIMF]{Pflamm2008}. These authors applied the IGIMF1 formulation and assumed the disk galaxy to be sub-divided into radial annular bins within each of which the local IGIMF is calculated subject to the constraint that the galaxy has an exponential radial structure and the gas and star formation rate densities are related.  
This work showed that the radial H$\alpha$ cut-off and extended UV disks can be explained naturally within the IGIMF framework because in the outskirts the gas density is low leading to a low star formation rate density, low-mass embedded clusters and thus a deficit of ionising stars, while intermediate-mass stars do form there. 
In addition, the cIGIMF results in metallicity gradients, as are suggested to be present for example in elliptical galaxies 
\citep{McConnell2016} and it provides a description for low surface-brightness galaxies (LSBGs) having low gas surface densities
\citep{Pflamm2011}. LSBGs, based on the cIGIMF calculation, form preferentially low mass stars with a deficit of high mass ones relative to the canonical IMF even though the global SFR can be high and the IGIMF would predict massive stars to be formed. We plan to include a mathematically and physically consistent cIGIMF description into the next version of the galIMF code originally developed by \cite{Yan2017}.

\subsection{Changes to the IMF variations within the IGIMF framework}
As formulated here, the IGIMF implements several empirical relations such as the star-mass function of embedded clusters, 
its variation, the correlation between the birth radius and mass of the embedded clusters,  and local IMF variations with the physical conditions in the star-forming cloud core. Since these are empirically derived not covering all possible physical values (extreme SFRs and metallicities are not accessible in the Local Universe for example), the IGIMF prescription applied here can be improved with time. That is, obtaining better data or data from so far not probed environments on a galactic scale and on larger scales can be used to infer local IMF variations. This has been shown here for the case of UFD measurements from \cite{Gennaro2018} (see Fig.~\ref{fig:local_cor} and Eq.~\ref{eq:alpha_cor}) in contrast to the original empirical extrapolation from MW data described by Eq.~\ref{eq:alpha12}.

Any proposed changes can be readily implemented into the galIMF code \citep{Yan2017} and further tested. However any local IMF variations need to match the canonical IMF for a SFR comparable to that of the Milky Way and an average MW metallicity as well as the present-day mass functions observed in globular clusters, open clusters and embedded clusters as a necessary constraint on any viable IMF theory.

\section{Conclusions}
For the first time a grid of galaxy-wide IMFs computed within the IGIMF framework with SFR and metallicity dependence is presented, together with 
its implementation into the galIMF module and an equivalent FORTRAN code. This allow us to trace the variations of galaxy-wide IMFs for different 
galaxies assuming that the physics driving the galaxy-wide IMFs comes from local star forming regions. 
The main contribution of this work can be summarized in a few points:
\begin{itemize}
    \item The attached IGIMF grid with the parameter-span SFR $\in (10^{-5},10^{4})$ and [Fe/H] $\in (-5,1)$ presents the
	galaxy-wide IMF normalised to the total stellar mass formed in~10~Myr episodes, $M_{tot} = \mathrm{SFR} \times 10\mathrm{Myr}$, always with the same range of masses of individual 
	stars (from 0.08 to 120 $M_{\odot}$) for an implementation into galaxy-evolution (e.g. chemo-dynamical evolution) codes and also for a possible interpolation in the grid. 
    \item The overall variation of the gwIMF is as follows: 
    (1) The gwIMF can become top-light even if the shape of the local stellar IMF is invariant (IGIMF1 version). The reason why this is the case can been seen 
    with a demonstrative example: 1000 star clusters with a mass of $10\, M_{\odot}$ would have a top-light stellar population in comparison to a monolithically formed star cluster of $10^4\, M_{\odot}$ because stars more massive than $10\, M_{\odot}$ would exist only in the latter case \citep{Yan2017}. This statement is basically independent of metallicity and reflects the fact that there is a maximum stellar mass that forms in a given cluster due to the $m_{\mathrm{max}}=WK(M_{\mathrm{ecl}})$ relation, and also that the upper limit for the most massive star cluster to be formed in a galaxy depends on galactic properties \citep{Johnson2017}. 
    The top-light IGIMF appears to be in good agreement with gwIMF measurements in nearby dwarf galaxies \citep{Lee2009, Watts2018}. The above demonstrative example is actually found in nature \citep{Hsu2012,Hsu2013}. 
    (2) The gwIMF, as expected, is close to the canonical IMF for a SFR near $1 \, \mathrm{M_{\odot}/yr}$ and Solar metallicity and becomes top-heavy with increasing SFR above that value.    
    (3) Interestingly, for sub-Solar metallicity the gwIMF can be bottom-light and for super-Solar metallicity bottom-heavy (IGIMF3 parametrization which includes the full metallicity and density variation of the stellar IMF). This might be reflected in the cores of elliptical galaxies.

\item We present the possible time evolution of the gwIMF for the case of a monolithic (star-burst)  formation of an elliptical galaxy showing the potential to explain observed features: high $\alpha$ element abundances implying a short formation time scale and high metallicity implying a top-heavy gwIMF during this short assembly time, 
and an over abundance of low-mass stars indicating  a metal-rich formation epoch in which the gwIMF is bottom-heavy. 

 \item As a consequence of the variable gwIMF the majority of standard and widely used stellar-population correlations used to estimate galaxy properties need to be recomputed and re-interpreted correctly. The reason is that into many of these, assumptions on the IMF enter. This is shown using the SFR--H$\alpha$ relation.  Assuming the IGIMF theory to be the correct description it follows that the SFR is underestimated for galaxies with a low SFR and it is overestimated for galaxies with high SFRs, by up to several orders of magnitude, if the standard Kennicutt SFR$_{\rm K}$--H$\alpha$ relation is used. We present the appropriate correction factors (Fig.~\ref{fig:sfr}). The Leo~P galaxy, often mentioned to pose a significant problem 
for the IGIMF theory, because for the estimated SFR stars as massive as $25\,M_\odot$ should not be forming in it, is shown to be well reproduced using the IGIMF theory. 
The reason is that, in the literature, the Kennicutt SFR$_{\rm K}$--H$\alpha$ relation
in incorreclty applied when star cluster masses are smaller than the most massive stars, i.e. when $M_{min} < m_{max}$. In this case the ensemble of freshly formed star clusters will contain low-mass clusters within which the IMF cannot be sampled to the most massive stars in the case of the IMF being a constrained probability distribution as in the SLUG approach (footnote~\ref{foot:SLUG}). This implies a systematic deficit of massive stars in the whole ensemble. This produces a biased result when the Kennicutt relation is extrapolated to low $H\alpha$ fluxes, that is when $SFR_K \lessapprox 10^{-4} M_{\odot}/\mathrm{yr}$ because the Kennicutt relation assumes all star-formation events to always have a fully sampled IMF \citep{daSilva2012}. This is naturally corrected for in the IGIMF theory such that the observed galaxy (e.g. Leo~P) behaves physically correctly (see Fig.\ref{fig:sfr}).

\item The data by \cite{Gennaro2018} for UFD galaxies suggest a possible variation of the low mass IMF for stars in the mass range (0.4-0.8 $M_{\odot}$) and metallicity in the range ([Fe/H]$=-3$ to $-2$). The IMF variations over this mass range and metallicity range have not been constrained empirically and only an extrapolation has been applied here (\citealt{Kroupa2001,Marks2012TH}, see Eq.~\ref{eq:alpha12}).  
The IGIMF theory is used to translate the observed galaxy-wide variation and to possibly improve the formulation of the local IMF variation in this stellar mass range and at low metallicity ([Fe/H] $= -3$ to $-2$). The newly suggested variation, see Eq.~\ref{eq:alpha_cor} and Fig.~\ref{fig:local_cor}, shows how gwIMF measurements can help constrain  star formation on star cluster scales, but we note the caveats discussed in Sec.~\ref{sec:UFDs}.
\end{itemize}

The IGIMF theory which is, by construction, consistent with Milky Way data, has now demonstrated its general potential in allowing the computations of gwIMFs from local empirical stellar IMF properties and also  the ability to improve these. It is ready to be implemented into chemo-dynamical codes and to be tested with more data.  First implementations of the IGIMF theory into self-consistent galaxy formation and evolution simulations have been achieved \citep{Bekki13, Ploeckinger2014}. The implications of the IGIMF theory for the gas-depletion and stellar-mass build-up time scales of galaxies are significant \citep{Pflamm2009b}, with many other galaxy-evolution problems being potentially resolved naturally \citep{Pflamm2011}. The SFR correction factors shown in Fig.~\ref{fig:sfr}, which need to be applied to the traditional Kennicutt values and in general entail the shift of low-mass-star-dominated populations in dwarf galaxies to massive-star-dominated populations in star-bursting galaxies, lead to a change of the slope of the galaxy main-sequence \citep{Speagle2014} 
which will be addressed in an upcoming contribution. Finally, the occurrence of H$\alpha$-dark star formation may significantly affect the cosmological SFR \citep{Pflamm2007}.

\begin{acknowledgements}
The authors acknowledge the exceptionally useful and comprehensive comments from an anonymous referee. We also thank to ESO visitor program for supporting AV and PK during their stay.
We thank Frederico Lelli and Glenn van de Ven for useful discussions and comments to the manuscript. AV acknowledges support from grant AYA2016-77237-C3-1-P from the Spanish Ministry of Economy and Competitiveness (MINECO).
AHZ acknowledges support from Alexander von Humboldt Foundation through a Research Fellowship. The discussions and the provided references by Neal Evans are also acknowledged.
\end{acknowledgements}

\bibliographystyle{aa}
\bibliography{library}

\begin{thebibliography}{162}
\expandafter\ifx\csname natexlab\endcsname\relax\def\natexlab#1{#1}\fi

\bibitem[{{Adams} \& {Fatuzzo}(1996)}]{AF96}
{Adams}, F.~C. \& {Fatuzzo}, M. 1996, \apj, 464, 256

\bibitem[{{Andr{\'e}} {et~al.}(2016){Andr{\'e}}, {Rev{\'e}ret}, {K{\"o}nyves},
  {Arzoumanian}, {Tig{\'e}}, {Gallais}, {Roussel}, {Le Pennec}, {Rodriguez},
  {Doumayrou}, {Dubreuil}, {Lortholary}, {Martignac}, {Talvard}, {Delisle},
  {Visticot}, {Dumaye}, {De Breuck}, {Shimajiri}, {Motte}, {Bontemps},
  {Hennemann}, {Zavagno}, {Russeil}, {Schneider}, {Palmeirim}, {Peretto},
  {Hill}, {Minier}, {Roy}, \& {Rygl}}]{Andre2016}
{Andr{\'e}}, P., {Rev{\'e}ret}, V., {K{\"o}nyves}, V., {et~al.} 2016, \aap,
  592, A54

\bibitem[{{Ascenso}(2018)}]{Ascenso2018}
{Ascenso}, J. 2018, in Astrophysics and Space Science Library, Vol. 424, The
  Birth of Star Clusters, ed. S.~{Stahler}, 1

\bibitem[{Ballesteros-Paredes {et~al.}(2007)Ballesteros-Paredes, Klessen,
  Mac~Low, \& Vazquez-Semadeni}]{Ballesteros2007}
Ballesteros-Paredes, J., Klessen, R.~S., Mac~Low, M.~M., \& Vazquez-Semadeni,
  E. 2007, Protostars and Planets V, 63

\bibitem[{Banerjee \& Kroupa(2013)}]{Banerjee2013}
Banerjee, S. \& Kroupa, P. 2013, \apj, 764, 29

\bibitem[{Banerjee \& Kroupa(2014)}]{Banerjee2014}
Banerjee, S. \& Kroupa, P. 2014, \apj, 787, 158

\bibitem[{Banerjee \& Kroupa(2015)}]{Banerjee2015}
Banerjee, S. \& Kroupa, P. 2015, \mnras, 447, 728

\bibitem[{Banerjee \& Kroupa(2017)}]{Banerjee2017}
Banerjee, S. \& Kroupa, P. 2017, \aap, 597, A28

\bibitem[{{Banerjee} {et~al.}(2012){Banerjee}, {Kroupa}, \& {Oh}}]{BKO12}
{Banerjee}, S., {Kroupa}, P., \& {Oh}, S. 2012, \mnras, 426, 1416

\bibitem[{{Bastian} {et~al.}(2010){Bastian}, {Covey}, \& {Meyer}}]{Bastian2010}
{Bastian}, N., {Covey}, K.~R., \& {Meyer}, M.~R. 2010, \araa, 48, 339

\bibitem[{Bate {et~al.}(2014)Bate, Tricco, \& Price}]{Bate2014}
Bate, M.~R., Tricco, T.~S., \& Price, D.~J. 2014, \mnras, 437, 77

\bibitem[{{Baumgardt} \& {Makino}(2003)}]{BaumgardtMakino2003}
{Baumgardt}, H. \& {Makino}, J. 2003, \mnras, 340, 227

\bibitem[{{Bekki}(2013)}]{Bekki13}
{Bekki}, K. 2013, \mnras, 436, 2254

\bibitem[{{Binney} \& {Tremaine}(1987)}]{BinneyTremaine1987}
{Binney}, J. \& {Tremaine}, S. 1987, {Galactic dynamics}

\bibitem[{{Boquien} {et~al.}(2014){Boquien}, {Buat}, \& {Perret}}]{Boquien2014}
{Boquien}, M., {Buat}, V., \& {Perret}, V. 2014, \aap, 571, A72

\bibitem[{Bressert {et~al.}(2010)Bressert, Bastian, Gutermuth, Megeath, Allen,
  Evans, Rebull, Hatchell, Johnstone, Bourke, Cieza, Harvey, Merin, Ray, \&
  Tothill}]{Bressert2010}
Bressert, E., Bastian, N., Gutermuth, R., {et~al.} 2010, \mnras: Letters, 409,
  L54

\bibitem[{Brinkmann {et~al.}(2017)Brinkmann, Banerjee, Motwani, \&
  Kroupa}]{Brinkmann2017}
Brinkmann, N., Banerjee, S., Motwani, B., \& Kroupa, P. 2017, \aap, 600, A49

\bibitem[{{Calzetti}(2013)}]{Calzeti2013}
{Calzetti}, D. 2013, {Star Formation Rate Indicators}, ed.
  J.~{Falc{\'o}n-Barroso} \& J.~H. {Knapen}, 419

\bibitem[{{Chabrier}(2003)}]{Chabrier2003}
{Chabrier}, G. 2003, \pasp, 115, 763

\bibitem[{{Conroy} {et~al.}(2017){Conroy}, {van Dokkum}, \&
  {Villaume}}]{Conroy2017}
{Conroy}, C., {van Dokkum}, P.~G., \& {Villaume}, A. 2017, \apj, 837, 166

\bibitem[{{da Silva} {et~al.}(2012){da Silva}, {Fumagalli}, \&
  {Krumholz}}]{daSilva2012}
{da Silva}, R.~L., {Fumagalli}, M., \& {Krumholz}, M. 2012, \apj, 745, 145

\bibitem[{da~Silva {et~al.}(2014)da~Silva, Fumagalli, \&
  Krumholz}]{daSilva2014}
da~Silva, R.~L., Fumagalli, M., \& Krumholz, M.~R. 2014, \mnras, 444, 3275

\bibitem[{{Dabringhausen} {et~al.}(2010){Dabringhausen}, {Fellhauer}, \&
  {Kroupa}}]{Dabringhausen2010}
{Dabringhausen}, J., {Fellhauer}, M., \& {Kroupa}, P. 2010, \mnras, 403, 1054

\bibitem[{Dabringhausen {et~al.}(2009)Dabringhausen, Kroupa, \&
  Baumgardt}]{Dabringhausen2009}
Dabringhausen, J., Kroupa, P., \& Baumgardt, H. 2009, \mnras, 394, 1529

\bibitem[{{Dabringhausen} {et~al.}(2012){Dabringhausen}, {Kroupa}, \&
  {Pflamm-Altenburg}}]{Dabringhausen2012}
{Dabringhausen}, J., {Kroupa}, P., \& {Pflamm-Altenburg}, J. and.~{Mieske}, S.
  2012, \apj, 747, 72

\bibitem[{{De Masi} {et~al.}(2018){De Masi}, {Matteucci}, \&
  {Vincenzo}}]{DeMasi2018a}
{De Masi}, C., {Matteucci}, F., \& {Vincenzo}, F. 2018, \mnras, 474, 5259

\bibitem[{Dobbs {et~al.}(2011)Dobbs, Burkert, \& Pringle}]{Dobbs2011}
Dobbs, C.~L., Burkert, A., \& Pringle, J.~E. 2011, \mnras, 413, 2935

\bibitem[{{Duarte-Cabral} {et~al.}(2013){Duarte-Cabral}, {Bontemps}, {Motte},
  {Hennemann}, {Schneider}, \& {Andr{\'e}}}]{Duarte13}
{Duarte-Cabral}, A., {Bontemps}, S., {Motte}, F., {et~al.} 2013, \aap, 558,
  A125

\bibitem[{Egusa {et~al.}(2009)Egusa, Kohno, Sofue, Nakanishi, \&
  Komugi}]{Egusa2009}
Egusa, F., Kohno, K., Sofue, Y., Nakanishi, H., \& Komugi, S. 2009, \apj, 697,
  1870

\bibitem[{Egusa {et~al.}(2004)Egusa, Sofue, \& Nakanishi}]{Egusa2004}
Egusa, F., Sofue, Y., \& Nakanishi, H. 2004, Publications of the Astronomical
  Society of Japan, 56, L45

\bibitem[{Elmegreen(2002)}]{Elmegreen2002}
Elmegreen, B.~G. 2002, \apj, 577, 206

\bibitem[{Elmegreen(2007)}]{Elmegreen2007}
Elmegreen, B.~G. 2007, \apj, 668, 1064

\bibitem[{{Famaey} \& {McGaugh}(2012)}]{FM12}
{Famaey}, B. \& {McGaugh}, S.~S. 2012, Living Reviews in Relativity, 15, 10

\bibitem[{Federrath(2015)}]{Federrath2015}
Federrath, C. 2015, \mnras, 450, 4035

\bibitem[{Federrath(2016)}]{Federrath2016}
Federrath, C. 2016, \mnras, 457, 375

\bibitem[{Federrath {et~al.}(2014)Federrath, Schr{\"o}n, Banerjee, \&
  Klessen}]{Federrath2014}
Federrath, C., Schr{\"o}n, M., Banerjee, R., \& Klessen, R.~S. 2014, \apj, 790,
  128

\bibitem[{{Ferreras} {et~al.}(2013){Ferreras}, {La Barbera}, {de la Rosa},
  {Vazdekis}, {de Carvalho}, {Falc{\'o}n-Barroso}, \&
  {Ricciardelli}}]{Ferreras+13}
{Ferreras}, I., {La Barbera}, F., {de la Rosa}, I.~G., {et~al.} 2013, \mnras,
  429, L15

\bibitem[{{Ferreras} {et~al.}(2015){Ferreras}, {Weidner}, {Vazdekis}, \& {La
  Barbera}}]{Ferreras+15}
{Ferreras}, I., {Weidner}, C., {Vazdekis}, A., \& {La Barbera}, F. 2015,
  \mnras, 448, L82

\bibitem[{{Figer}(2005)}]{Figer05}
{Figer}, D.~F. 2005, \nat, 434, 192

\bibitem[{{Fioc} {et~al.}(2011){Fioc}, {Le Borgne}, \&
  {Rocca-Volmerange}}]{Fioc2011}
{Fioc}, M., {Le Borgne}, D., \& {Rocca-Volmerange}, B. 2011, {P{\'E}GASE:
  Metallicity-consistent Spectral Evolution Model of Galaxies}, Astrophysics
  Source Code Library

\bibitem[{{Fioc} \& {Rocca-Volmerange}(1999)}]{Fioc1999}
{Fioc}, M. \& {Rocca-Volmerange}, B. 1999, ArXiv Astrophysics e-prints, code
  description, only astro-ph version

\bibitem[{{Fontanot} {et~al.}(2017){Fontanot}, {De Lucia}, {Hirschmann},
  {Bruzual}, \& {Zibetti}}]{Fontanot2017}
{Fontanot}, F., {De Lucia}, G., {Hirschmann}, M., {Bruzual}, G. and.~{Charlot},
  S., \& {Zibetti}, S. 2017, \mnras, 464, 3812

\bibitem[{{Fontanot} {et~al.}(2018{\natexlab{a}}){Fontanot}, {De Lucia}, {Xie},
  {Hirschmann}, {Bruzual}, \& {Charlot}}]{Fontanot2018b}
{Fontanot}, F., {De Lucia}, G., {Xie}, L., {et~al.} 2018{\natexlab{a}}, \mnras,
  475, 2467

\bibitem[{{Fontanot} {et~al.}(2018{\natexlab{b}}){Fontanot}, {La Barbera}, {De
  Lucia}, {Pasquali}, \& {Vazdekis}}]{Fontanot2018a}
{Fontanot}, F., {La Barbera}, F., {De Lucia}, G., {Pasquali}, A., \&
  {Vazdekis}, A. 2018{\natexlab{b}}, \mnras, 479, 5678

\bibitem[{Fukui \& Kawamura(2010)}]{Fukui2010}
Fukui, Y. \& Kawamura, A. 2010, \araa, 48, 547

\bibitem[{{Geha} {et~al.}(2013){Geha}, {Brown}, {Tumlinson}, {Kalirai},
  {Simon}, {Kirby}, {VandenBerg}, {Mu{\~n}oz}, {Avila}, {Guhathakurta}, \&
  {Ferguson}}]{Geha2013}
{Geha}, M., {Brown}, T.~M., {Tumlinson}, J., {et~al.} 2013, \apj, 771, 29

\bibitem[{{Gennaro} {et~al.}(2018{\natexlab{a}}){Gennaro}, {Geha},
  {Tchernyshyov}, {Brown}, {Avila}, {Conroy}, {Mu{\~n}oz}, {Simon}, \&
  {Tumlinson}}]{Gennaro2018b}
{Gennaro}, M., {Geha}, M., {Tchernyshyov}, K., {et~al.} 2018{\natexlab{a}},
  \apj, 863, 38

\bibitem[{{Gennaro} {et~al.}(2018{\natexlab{b}}){Gennaro}, {Tchernyshyov},
  {Brown}, {Geha}, {Avila}, {Guhathakurta}, {Kalirai}, {Kirby}, {Renzini},
  {Simon}, {Tumlinson}, \& {Vargas}}]{Gennaro2018}
{Gennaro}, M., {Tchernyshyov}, K., {Brown}, T.~M., {et~al.} 2018{\natexlab{b}},
  \apj, 855, 20

\bibitem[{{Gieles} {et~al.}(2012){Gieles}, {Moeckel}, \& {Clarke}}]{Gieles2012}
{Gieles}, M., {Moeckel}, N., \& {Clarke}, C.~J. 2012, \mnras, 426, L11

\bibitem[{Giovanelli {et~al.}(2013)Giovanelli, Haynes, Adams, Cannon, Rhode,
  Salzer, Skillman, Bernstein-Cooper, \& McQuinn}]{Giovanelli2013}
Giovanelli, R., Haynes, M.~P., Adams, E. A.~K., {et~al.} 2013, \aj, 146, 15

\bibitem[{{Gunawardhana} {et~al.}(2011){Gunawardhana}, {Hopkins}, {Sharp},
  {Taylor}, {Bland-Hawthorn}, {Maraston}, {Popescu}, {Wijesinghe}, {Croom},
  {Sadler}, {Wilkins}, {Liske}, {Norberg}, {Baldry}, {Loveday}, {Peacock},
  {Zucker}, {Parker}, {Cameron}, {Frenk}, {Kelvin}, {Kuijken}, {Madore},
  {Parkinson}, {Pimbblet}, {Sutherland}, \& {Thomas}}]{Gunawardhana2011}
{Gunawardhana}, M.~L.~P., {Hopkins}, A.~M., {Sharp}, R.~G. and.~{Brough}, S.,
  {et~al.} 2011, \mnras, 415, 1647

\bibitem[{{Hacar} {et~al.}(2017{\natexlab{a}}){Hacar}, {Alves}, {Tafalla}, \&
  {Goicoechea}}]{Hacar+17}
{Hacar}, A., {Alves}, J., {Tafalla}, M., \& {Goicoechea}, J.~R.
  2017{\natexlab{a}}, \aap, 602, L2

\bibitem[{{Hacar} {et~al.}(2017{\natexlab{b}}){Hacar}, {Tafalla}, \&
  {Alves}}]{Hacar+17b}
{Hacar}, A., {Tafalla}, M., \& {Alves}, J. 2017{\natexlab{b}}, \aap, 606, A123

\bibitem[{Haghi {et~al.}(2017)Haghi, Khalaj, Hasani~Zonoozi, \&
  Kroupa}]{Haghi2017}
Haghi, H., Khalaj, P., Hasani~Zonoozi, A., \& Kroupa, P. 2017, \apj, 839, 60

\bibitem[{{Haghi} {et~al.}(2015){Haghi}, {Zonoozi}, {Kroupa}, {Banerjee}, \&
  {Baumgardt}}]{Haghi2015}
{Haghi}, H., {Zonoozi}, A.~H., {Kroupa}, P., {Banerjee}, S., \& {Baumgardt}, H.
  2015, \mnras, 454, 3872

\bibitem[{{Hansen} {et~al.}(2012){Hansen}, {Klein}, {McKee}, \&
  {Fisher}}]{Hansen2012}
{Hansen}, C.~E., {Klein}, R.~I., {McKee}, C.~F., \& {Fisher}, R.~T. 2012, \apj,
  747, 22

\bibitem[{Hartmann {et~al.}(2001)Hartmann, Ballesteros-Paredes, \&
  Bergin}]{Hartmann2001}
Hartmann, L., Ballesteros-Paredes, J., \& Bergin, E.~A. 2001, \apj, 562, 852

\bibitem[{{Heggie} \& {Hut}(2003)}]{HeggieHut2003}
{Heggie}, D. \& {Hut}, P. 2003, {The Gravitational Million-Body Problem: A
  Multidisciplinary Approach to Star Cluster Dynamics}

\bibitem[{{Hopkins}(2018)}]{Hopkins2018}
{Hopkins}, A.~M. 2018, ArXiv e-prints, accepted for publication in PASA

\bibitem[{Hsu {et~al.}(2012)Hsu, Hartmann, Allen, Hern{\'a}ndez, Megeath,
  Mosby, Tobin, \& Espaillat}]{Hsu2012}
Hsu, W.-H., Hartmann, L., Allen, L., {et~al.} 2012, \apj, 752, 59

\bibitem[{Hsu {et~al.}(2013)Hsu, Hartmann, Allen, Hern{\'a}ndez, Megeath,
  Tobin, \& Ingleby}]{Hsu2013}
Hsu, W.-H., Hartmann, L., Allen, L., {et~al.} 2013, \apj, 764, 114

\bibitem[{Je{\v{r}}{\'a}bkov{\'a} {et~al.}(2017)Je{\v{r}}{\'a}bkov{\'a},
  Kroupa, Dabringhausen, Hilker, \& Bekki}]{Jerabkova2017}
Je{\v{r}}{\'a}bkov{\'a}, T., Kroupa, P., Dabringhausen, J., Hilker, M., \&
  Bekki, K. 2017, \aap, 608, A53

\bibitem[{Johnson {et~al.}(2017)Johnson, Seth, Dalcanton, Beerman, Fouesneau,
  Weisz, Bell, Dolphin, Sandstrom, \& Williams}]{Johnson2017}
Johnson, L.~C., Seth, A.~C., Dalcanton, J.~J., {et~al.} 2017, \apj, 839, 78

\bibitem[{{Kalari} {et~al.}(2018){Kalari}, {Carraro}, {Evans}, \&
  {Rubio}}]{Kalari2018}
{Kalari}, V.~M., {Carraro}, G., {Evans}, C.~J., \& {Rubio}, M. 2018, \apj, 857,
  132

\bibitem[{Kennicutt(1998)}]{Kennicutt1998}
Kennicutt, R. C.~J. 1998, \araa, 36, 189

\bibitem[{{Kirk} \& {Myers}(2011)}]{Kirk2011}
{Kirk}, H. \& {Myers}, P.~C. 2011, \apj, 727, 64

\bibitem[{{Kirk} \& {Myers}(2012)}]{Kirk2012}
{Kirk}, H. \& {Myers}, P.~C. 2012, \apj, 745, 131

\bibitem[{{Koen}(2006)}]{Koen06}
{Koen}, C. 2006, \mnras, 365, 590

\bibitem[{{Kroupa}(2001)}]{Kroupa2001}
{Kroupa}, P. 2001, \mnras, 322, 231

\bibitem[{{Kroupa}(2002)}]{Kroupa2002}
{Kroupa}, P. 2002, Science, 295, 82

\bibitem[{{Kroupa}(2005)}]{Kroupa2005}
{Kroupa}, P. 2005, in ESA Special Publication, Vol. 576, The Three-Dimensional
  Universe with Gaia, ed. C.~{Turon}, K.~S. {O'Flaherty}, \& M.~A.~C.
  {Perryman}, 629

\bibitem[{{Kroupa}(2015)}]{Kroupa2015}
{Kroupa}, P. 2015, Canadian Journal of Physics, 93, 169

\bibitem[{Kroupa {et~al.}(2001)Kroupa, Aarseth, \& Hurley}]{KAH}
Kroupa, P., Aarseth, S., \& Hurley, J. 2001, \mnras, 321, 699

\bibitem[{{Kroupa} \& {Boily}(2002)}]{KroupaBoily2002}
{Kroupa}, P. \& {Boily}, C.~M. 2002, \mnras, 336, 1188

\bibitem[{{Kroupa} \& {Bouvier}(2003)}]{Kroupa2003}
{Kroupa}, P. \& {Bouvier}, J. 2003, \mnras, 346, 343

\bibitem[{{Kroupa} \& {Jerabkova}(2018)}]{KJ2018}
{Kroupa}, P. \& {Jerabkova}, T. 2018, ArXiv e-prints

\bibitem[{{Kroupa} {et~al.}(2018){Kroupa}, {Je{\v r}{\'a}bkov{\'a}},
  {Dinnbier}, {Beccari}, \& {Yan}}]{Kroupa2018}
{Kroupa}, P., {Je{\v r}{\'a}bkov{\'a}}, T., {Dinnbier}, F., {Beccari}, G., \&
  {Yan}, Z. 2018, \aap, 612, A74

\bibitem[{{Kroupa} {et~al.}(1993){Kroupa}, {Tout}, \& {Gilmore}}]{Kroupa1993}
{Kroupa}, P., {Tout}, C.~A., \& {Gilmore}, G. 1993, \mnras, 262, 545

\bibitem[{{Kroupa} \& {Weidner}(2003)}]{KroupaWeidner2003}
{Kroupa}, P. \& {Weidner}, C. 2003, \apj, 598, 1076

\bibitem[{{Kroupa} {et~al.}(2013){Kroupa}, {Weidner}, {Pflamm-Altenburg},
  {Dabringhausen}, {Marks}, \& {Maschberger}}]{Kroupa2013}
{Kroupa}, P., {Weidner}, C., {Pflamm-Altenburg}, J. and.~{Thies}, I., {et~al.}
  2013, {The Stellar and Sub-Stellar Initial Mass Function of Simple and
  Composite Populations} (Springer), 115

\bibitem[{{La Barbera} {et~al.}(2013){La Barbera}, {Ferreras}, {Vazdekis}, {de
  la Rosa}, {de Carvalho}, {Trevisan}, {Falc{\'o}n-Barroso}, \&
  {Ricciardelli}}]{LaBarbera+13}
{La Barbera}, F., {Ferreras}, I., {Vazdekis}, A., {et~al.} 2013, \mnras, 433,
  3017

\bibitem[{{Lada}(2010)}]{Lada2010}
{Lada}, C.~J. 2010, Philosophical Transactions of the Royal Society of London
  Series A, 368, 713

\bibitem[{{Lada} \& {Lada}(2003)}]{Lada2003}
{Lada}, C.~J. \& {Lada}, E.~A. 2003, \araa, 41, 57

\bibitem[{{Lee} {et~al.}(2009){Lee}, {Gil de Paz}, {Tremonti}, {Kennicutt},
  {Bothwell}, {Calzetti}, {Dalcanton}, {Engelbracht}, {Funes}, {Sakai},
  {Skillman}, {van Zee}, \& {Weisz}}]{Lee2009}
{Lee}, J.~C., {Gil de Paz}, A., {Tremonti}, C., {et~al.} 2009, \apj, 706, 599

\bibitem[{Lee {et~al.}(2002)Lee, Salzer, Impey, Thuan, \& Gronwall}]{Lee2002}
Lee, J.~C., Salzer, J.~J., Impey, C., Thuan, T.~X., \& Gronwall, C. 2002, \aj,
  124, 3088

\bibitem[{Leisawitz(1989)}]{Leisawitz1989}
Leisawitz, D. 1989, \baas, 21, 1189

\bibitem[{{Lelli} {et~al.}(2016){Lelli}, {McGaugh}, \& {Schombert}}]{Lelli+16}
{Lelli}, F., {McGaugh}, S.~S., \& {Schombert}, J.~M. 2016, \apjl, 816, L14

\bibitem[{Lieberz \& Kroupa(2017)}]{Lieberz2017}
Lieberz, P. \& Kroupa, P. 2017, \mnras, 465, 3775

\bibitem[{{Lim} {et~al.}(2018){Lim}, {Sung}, {Bessell}, {Lee}, {Lee}, {Oh},
  {Hwang}, {Park}, {Hur}, {Hong}, \& {Park}}]{Lim+18}
{Lim}, B., {Sung}, H., {Bessell}, M.~S., {et~al.} 2018, \mnras, 477, 1993

\bibitem[{{Lu} {et~al.}(2018){Lu}, {Zhang}, {Liu}, {Sanhueza}, {Tatematsu},
  {Feng}, {Smith}, {Myers}, {Sridharan}, \& {Gu}}]{Lu2018}
{Lu}, X., {Zhang}, Q., {Liu}, H.~B., {et~al.} 2018, \apj, 855, 9

\bibitem[{{Lucas} {et~al.}(2018){Lucas}, {Rybak}, {Bonnell}, \&
  {Gieles}}]{Lucas2018}
{Lucas}, W.~E., {Rybak}, M., {Bonnell}, I.~A., \& {Gieles}, M. 2018, \mnras,
  474, 3582

\bibitem[{Machida \& Matsumoto(2012)}]{Machida2012}
Machida, M.~N. \& Matsumoto, T. 2012, \mnras, 421, 588

\bibitem[{{Ma{\'{\i}}z Apell{\'a}niz} {et~al.}(2007){Ma{\'{\i}}z
  Apell{\'a}niz}, {Walborn}, {Morrell}, {Niemela}, \& {Nelan}}]{Maiz+2007}
{Ma{\'{\i}}z Apell{\'a}niz}, J., {Walborn}, N.~R., {Morrell}, N.~I., {Niemela},
  V.~S., \& {Nelan}, E.~P. 2007, \apj, 660, 1480

\bibitem[{{Marks} \& {Kroupa}(2012)}]{Marks2012}
{Marks}, M. \& {Kroupa}, P. 2012, \aap, 543, A8

\bibitem[{{Marks} {et~al.}(2012){Marks}, {Kroupa}, {Dabringhausen}, \&
  {Pawlowski}}]{Marks2012TH}
{Marks}, M., {Kroupa}, P., {Dabringhausen}, J., \& {Pawlowski}, M.~S. 2012,
  \mnras, 422, 2246

\bibitem[{{Marks} {et~al.}(2014){Marks}, {Kroupa}, {Dabringhausen}, \&
  {Pawlowski}}]{Marks2014_erratum}
{Marks}, M., {Kroupa}, P., {Dabringhausen}, J., \& {Pawlowski}, M.~S. 2014,
  \mnras, 442, 3315

\bibitem[{{Mart{\'{\i}}n-Navarro}(2016)}]{MartinNavarro16}
{Mart{\'{\i}}n-Navarro}, I. 2016, \mnras, 456, L104

\bibitem[{Massey(2003)}]{Massey2003}
Massey, P. 2003, \araa, 41, 15

\bibitem[{{Matteucci}(1994)}]{Matteucci1994}
{Matteucci}, F. 1994, \aap, 288, 57

\bibitem[{{McConnell} {et~al.}(2016){McConnell}, {Lu}, \&
  {Mann}}]{McConnell2016}
{McConnell}, N.~J., {Lu}, J.~R., \& {Mann}, A.~W. 2016, \apj, 821, 39

\bibitem[{{McGaugh} {et~al.}(2000){McGaugh}, {Schombert}, {Bothun}, \& {de
  Blok}}]{McGaugh00}
{McGaugh}, S.~S., {Schombert}, J.~M., {Bothun}, G.~D., \& {de Blok}, W.~J.~G.
  2000, \apjl, 533, L99

\bibitem[{McQuinn {et~al.}(2015)McQuinn, Skillman, Dolphin, Cannon, Salzer,
  Rhode, Adams, Berg, Giovanelli, Girardi, \& Haynes}]{McQuinn2015}
McQuinn, K. B.~W., Skillman, E.~D., Dolphin, A., {et~al.} 2015, \apj, 812, 158

\bibitem[{{Megeath} {et~al.}(2016){Megeath}, {Gutermuth}, {Muzerolle},
  {Kryukova}, {Hora}, {Allen}, {Flaherty}, {Hartmann}, {Myers}, {Pipher},
  {Stauffer}, {Young}, \& {Fazio}}]{Megeath2016}
{Megeath}, S.~T., {Gutermuth}, R., {Muzerolle}, J., {et~al.} 2016, \aj, 151, 5

\bibitem[{Meidt {et~al.}(2015)Meidt, Hughes, Dobbs, Pety, Thompson,
  Garc{\'\i}a-Burillo, Leroy, Schinnerer, Colombo, Querejeta, Kramer, Schuster,
  \& Dumas}]{Meidt2015}
Meidt, S.~E., Hughes, A., Dobbs, C.~L., {et~al.} 2015, \apj, 806, 72

\bibitem[{{Meurer} {et~al.}(2009){Meurer}, {Wong}, {Kim}, {Hanish}, {Werk},
  {Bland-Hawthorn}, {Zwaan}, {Koribalski}, {Thilker}, {Ferguson}, {Putman},
  {Knezek}, {Drinkwater}, {Hoopes}, {Meyer}, {Ryan-Weber}, \&
  {Staveley-Smith}}]{Meurer2009}
{Meurer}, G.~R., {Wong}, O.~I., {Kim}, J.~H., {et~al.} 2009, \apj, 695, 765

\bibitem[{{Milgrom}(1983)}]{Milgrom1983}
{Milgrom}, M. 1983, \apj, 270, 365

\bibitem[{{Neuh{\"a}user} {et~al.}(1998){Neuh{\"a}user}, {Frink}, {R{\"o}ser},
  {Torres}, {Brandner}, {Alcal{\'a}}, {Covino}, {Wichmann}, \&
  {Kunkel}}]{Neuhaueser1998}
{Neuh{\"a}user}, R., {Frink}, S., {R{\"o}ser}, S.~R., {et~al.} 1998, Acta
  Historica Astronomiae, 3, 206

\bibitem[{{Oey} \& {Clarke}(2005)}]{OeyClarke05}
{Oey}, M.~S. \& {Clarke}, C.~J. 2005, \apjl, 620, L43

\bibitem[{{Offner} {et~al.}(2014){Offner}, {Clark}, {Hennebelle}, {Bastian},
  {Bate}, {Hopkins}, {Moraux}, \& {Whitworth}}]{Offner2014}
{Offner}, S.~S.~R., {Clark}, P.~C., {Hennebelle}, P., {et~al.} 2014, Protostars
  and Planets VI, 53

\bibitem[{{Oh} \& {Kroupa}(2012)}]{OK12}
{Oh}, S. \& {Kroupa}, P. 2012, \mnras, 424, 65

\bibitem[{{Oh} \& {Kroupa}(2016{\natexlab{a}})}]{Oh2016}
{Oh}, S. \& {Kroupa}, P. 2016{\natexlab{a}}, \aap, 590, A107

\bibitem[{{Oh} \& {Kroupa}(2016{\natexlab{b}})}]{OK16}
{Oh}, S. \& {Kroupa}, P. 2016{\natexlab{b}}, \aap, 590, A107

\bibitem[{{Oh} {et~al.}(2015{\natexlab{a}}){Oh}, {Kroupa}, \&
  {Pflamm-Altenburg}}]{Oh2015}
{Oh}, S., {Kroupa}, P., \& {Pflamm-Altenburg}, J. 2015{\natexlab{a}}, \apj,
  805, 92

\bibitem[{{Oh} {et~al.}(2015{\natexlab{b}}){Oh}, {Kroupa}, \&
  {Pflamm-Altenburg}}]{Oh+15}
{Oh}, S., {Kroupa}, P., \& {Pflamm-Altenburg}, J. 2015{\natexlab{b}}, \apj,
  805, 92

\bibitem[{{Padoan} {et~al.}(2017){Padoan}, {Haugb{\o}lle}, {Nordlund}, \&
  {Frimann}}]{Padoan2017}
{Padoan}, P., {Haugb{\o}lle}, T., {Nordlund}, {\AA}., \& {Frimann}, S. 2017,
  \apj, 840, 48

\bibitem[{{Padoan} {et~al.}(2016){Padoan}, {Pan}, {Haugb{\o}lle}, \&
  {Nordlund}}]{Padoan2016}
{Padoan}, P., {Pan}, L., {Haugb{\o}lle}, T., \& {Nordlund}, {\AA}. 2016, \apj,
  822, 11

\bibitem[{{Papadopoulos}(2010)}]{Papadopoulos2010}
{Papadopoulos}, P.~P. 2010, \apj, 720, 226

\bibitem[{{Pflamm-Altenburg} {et~al.}(2013){Pflamm-Altenburg},
  {Gonz{\'a}lez-L{\'o}pezlira}, \& {Kroupa}}]{Pflamm2013}
{Pflamm-Altenburg}, J., {Gonz{\'a}lez-L{\'o}pezlira}, R.~A., \& {Kroupa}, P.
  2013, \mnras, 435, 2604

\bibitem[{{Pflamm-Altenburg} \& {Kroupa}(2008{\natexlab{a}})}]{Pflamm2008}
{Pflamm-Altenburg}, J. \& {Kroupa}, P. 2008{\natexlab{a}}, \nat, 455, 641

\bibitem[{{Pflamm-Altenburg} \&
  {Kroupa}(2008{\natexlab{b}})}]{PflammKroupa2008}
{Pflamm-Altenburg}, J. \& {Kroupa}, P. 2008{\natexlab{b}}, \nat, 455, 641

\bibitem[{{Pflamm-Altenburg} \& {Kroupa}(2009)}]{Pflamm2009b}
{Pflamm-Altenburg}, J. \& {Kroupa}, P. 2009, \apj, 706, 516

\bibitem[{Pflamm-Altenburg {et~al.}(2007)Pflamm-Altenburg, Weidner, \&
  Kroupa}]{Pflamm2007}
Pflamm-Altenburg, J., Weidner, C., \& Kroupa, P. 2007, \apj, 671, 1550

\bibitem[{{Pflamm-Altenburg} {et~al.}(2009){Pflamm-Altenburg}, {Weidner}, \&
  {Kroupa}}]{Pflamm2009}
{Pflamm-Altenburg}, J., {Weidner}, C., \& {Kroupa}, P. 2009, \mnras, 395, 394

\bibitem[{{Pflamm-Altenburg} {et~al.}(2011){Pflamm-Altenburg}, {Weidner}, \&
  {Kroupa}}]{Pflamm2011}
{Pflamm-Altenburg}, J., {Weidner}, C., \& {Kroupa}, P. 2011, in Astronomical
  Society of the Pacific Conference Series, Vol. 440, UP2010: Have Observations
  Revealed a Variable Upper End of the Initial Mass Function?, ed. M.~{Treyer},
  T.~{Wyder}, J.~{Neill}, M.~{Seibert}, \& J.~{Lee}, 269

\bibitem[{{Ploeckinger} {et~al.}(2014){Ploeckinger}, {Hensler}, {Recchi},
  {Mitchell}, \& {Kroupa}}]{Ploeckinger2014}
{Ploeckinger}, S., {Hensler}, G., {Recchi}, S., {Mitchell}, N., \& {Kroupa}, P.
  2014, \mnras, 437, 3980

\bibitem[{{Plunkett} {et~al.}(2018){Plunkett}, {Fern{\'a}ndez-L{\'o}pez},
  {Arce}, {Busquet}, {Mardones}, \& {Dunham}}]{Plunkett2018}
{Plunkett}, A.~L., {Fern{\'a}ndez-L{\'o}pez}, M., {Arce}, H.~G., {et~al.} 2018,
  \aap, 615, A9

\bibitem[{{Qiu} {et~al.}(2007){Qiu}, {Zhang}, {Beuther}, \& {Yang}}]{Qiu2007}
{Qiu}, K., {Zhang}, Q., {Beuther}, H., \& {Yang}, J. 2007, \apj, 654, 361

\bibitem[{{Qiu} {et~al.}(2008){Qiu}, {Zhang}, {Megeath}, {Gutermuth},
  {Beuther}, {Shepherd}, {Sridharan}, {Testi}, \& {De Pree}}]{Qiu2008}
{Qiu}, K., {Zhang}, Q., {Megeath}, S.~T., {et~al.} 2008, \apj, 685, 1005

\bibitem[{{Qiu} {et~al.}(2011){Qiu}, {Zhang}, \& {Menten}}]{Qiu2011}
{Qiu}, K., {Zhang}, Q., \& {Menten}, K.~M. 2011, \apj, 728, 6

\bibitem[{{Ram{\'{\i}}rez Alegr{\'{\i}}a} {et~al.}(2016){Ram{\'{\i}}rez
  Alegr{\'{\i}}a}, {Borissova}, {Chen{\'e}}, {Bonatto}, {Kurtev}, {Amigo},
  {Kuhn}, {Gromadzki}, \& {Carballo-Bello}}]{Ramirez+16}
{Ram{\'{\i}}rez Alegr{\'{\i}}a}, S., {Borissova}, J., {Chen{\'e}}, A.-N.,
  {et~al.} 2016, \aap, 588, A40

\bibitem[{{Recchi} {et~al.}(2009){Recchi}, {Calura}, \& {Kroupa}}]{Recchi2009}
{Recchi}, S., {Calura}, F., \& {Kroupa}, P. 2009, \aap, 499, 711

\bibitem[{{Recchi} \& {Kroupa}(2015)}]{Recchi2015}
{Recchi}, S. \& {Kroupa}, P. 2015, \mnras, 446, 4168

\bibitem[{Renaud {et~al.}(2016)Renaud, Famaey, \& Kroupa}]{Renaud2016}
Renaud, F., Famaey, B., \& Kroupa, P. 2016, \mnras, 463, 3637

\bibitem[{{Romano} {et~al.}(2017){Romano}, {Matteucci}, {Zhang},
  {Papadopoulos}, \& {Ivison}}]{Romano2017}
{Romano}, D., {Matteucci}, F., {Zhang}, Z.-Y., {Papadopoulos}, P.~P., \&
  {Ivison}, R.~J. 2017, \mnras, 470, 401

\bibitem[{{Salpeter}(1955)}]{Salpeter1955}
{Salpeter}, E.~E. 1955, \apj, 121, 161

\bibitem[{{Schneider} {et~al.}(2018){Schneider}, {Sana}, {Evans},
  {Bestenlehner}, {Castro}, {Fossati}, {Gr{\"a}fener}, {Langer},
  {Ram{\'{\i}}rez-Agudelo}, {Sab{\'{\i}}n-Sanjuli{\'a}n},
  {Sim{\'o}n-D{\'{\i}}az}, {Tramper}, {Crowther}, {de Koter}, {de Mink},
  {Dufton}, {Garcia}, {Gieles}, {H{\'e}nault-Brunet}, {Herrero}, {Izzard},
  {Kalari}, {Lennon}, {Ma{\'{\i}}z Apell{\'a}niz}, {Markova}, {Najarro},
  {Podsiadlowski}, {Puls}, {Taylor}, {van Loon}, {Vink}, \&
  {Norman}}]{Schneider2018}
{Schneider}, F.~R.~N., {Sana}, H., {Evans}, C.~J., {et~al.} 2018, Science, 359,
  69

\bibitem[{Schulz {et~al.}(2015)Schulz, Pflamm-Altenburg, \&
  Kroupa}]{Schulz2015}
Schulz, C., Pflamm-Altenburg, J., \& Kroupa, P. 2015, \aap, 582, A93

\bibitem[{{Selman} \& {Melnick}(2008)}]{Selman2008}
{Selman}, F.~J. \& {Melnick}, J. 2008, \apj, 689, 816

\bibitem[{{Smith} {et~al.}(2005){Smith}, {Stassun}, \& {Bally}}]{Smith2005}
{Smith}, N., {Stassun}, K.~G., \& {Bally}, J. 2005, \aj, 129, 888

\bibitem[{{Speagle} {et~al.}(2014){Speagle}, {Steinhardt}, {Capak}, \&
  {Silverman}}]{Speagle2014}
{Speagle}, J.~S., {Steinhardt}, C.~L., {Capak}, P.~L., \& {Silverman}, J.~D.
  2014, \apjs, 214, 15

\bibitem[{Stephens {et~al.}(2017)Stephens, Gouliermis, Looney, Gruendl, Chu,
  Weisz, Seale, Chen, Wong, Hughes, Pineda, Ott, \& Muller}]{Stephens2017}
Stephens, I.~W., Gouliermis, D., Looney, L.~W., {et~al.} 2017, \apj, 834, 94

\bibitem[{{Tafalla} {et~al.}(2002){Tafalla}, {Myers}, {Caselli}, {Walmsley}, \&
  {Comito}}]{Tafalla2002}
{Tafalla}, M., {Myers}, P.~C., {Caselli}, P., {Walmsley}, C.~M., \& {Comito},
  C. 2002, \apj, 569, 815

\bibitem[{{van Dokkum} \& {Conroy}(2010)}]{vanDokkum2010}
{van Dokkum}, P.~G. \& {Conroy}, C. 2010, \nat, 468, 940

\bibitem[{{Vanbeveren}(1982)}]{vanBeveren1982}
{Vanbeveren}, D. 1982, \aap, 115, 65

\bibitem[{{Vazdekis} {et~al.}(1997){Vazdekis}, {Peletier}, \&
  {Beckman}}]{Vazdekis1997}
{Vazdekis}, A., {Peletier}, R.~F., \& {Beckman}, J.~E. and.~{Casuso}, E. 1997,
  \apjs, 111, 203

\bibitem[{{Watts} {et~al.}(2018){Watts}, {Meurer}, {Lagos}, {Bruzzese},
  {Kroupa}, \& {Jerabkova}}]{Watts2018}
{Watts}, A.~B., {Meurer}, G.~R., {Lagos}, C.~D.~P., {et~al.} 2018, \mnras, 477,
  5554, in press

\bibitem[{{Weidner} {et~al.}(2013{\natexlab{a}}){Weidner}, {Ferreras},
  {Vazdekis}, \& {La Barbera}}]{Weidner2013c}
{Weidner}, C., {Ferreras}, I., {Vazdekis}, A., \& {La Barbera}, F.
  2013{\natexlab{a}}, \mnras, 435, 2274

\bibitem[{{Weidner} \& {Kroupa}(2004)}]{WeidnerKroupa2004}
{Weidner}, C. \& {Kroupa}, P. 2004, \mnras, 348, 187

\bibitem[{{Weidner} \& {Kroupa}(2005)}]{Weidner2005}
{Weidner}, C. \& {Kroupa}, P. 2005, \apj, 625, 754

\bibitem[{{Weidner} \& {Kroupa}(2006)}]{Weidner2006}
{Weidner}, C. \& {Kroupa}, P. 2006, \mnras, 365, 1333

\bibitem[{{Weidner} {et~al.}(2010){Weidner}, {Kroupa}, \&
  {Bonnell}}]{Weidner2010}
{Weidner}, C., {Kroupa}, P., \& {Bonnell}, I.~A.~D. 2010, \mnras, 401, 275

\bibitem[{{Weidner} {et~al.}(2004){Weidner}, {Kroupa}, \&
  {Larsen}}]{Weidner2004}
{Weidner}, C., {Kroupa}, P., \& {Larsen}, S.~S. 2004, \mnras, 350, 1503

\bibitem[{{Weidner} {et~al.}(2013{\natexlab{b}}){Weidner}, {Kroupa}, \&
  {Pflamm-Altenburg}}]{Weidner2013}
{Weidner}, C., {Kroupa}, P., \& {Pflamm-Altenburg}, J. and.~{Vazdekis}, A.
  2013{\natexlab{b}}, \mnras, 436, 3309

\bibitem[{{Weidner} {et~al.}(2013{\natexlab{c}}){Weidner}, {Kroupa}, \&
  {Pflamm-Altenburg}}]{Weidner2013b}
{Weidner}, C., {Kroupa}, P., \& {Pflamm-Altenburg}, J. 2013{\natexlab{c}},
  \mnras, 434, 84

\bibitem[{{Whitmore} {et~al.}(1999){Whitmore}, {Zhang}, {Leitherer}, {Fall},
  {Schweizer}, \& {Miller}}]{Whitmore1999}
{Whitmore}, B.~C., {Zhang}, Q., {Leitherer}, C., {et~al.} 1999, \aj, 118, 1551

\bibitem[{{Wright} \& {Mamajek}(2018)}]{Wright2018}
{Wright}, N.~J. \& {Mamajek}, E.~E. 2018, \mnras, 476, 381

\bibitem[{{Wu} {et~al.}(2010){Wu}, {Evans}, {Shirley}, \& {Knez}}]{Wu2010}
{Wu}, J., {Evans}, II, N.~J., {Shirley}, Y.~L., \& {Knez}, C. 2010, \apjs, 188,
  313

\bibitem[{{Wuchterl} \& {Tscharnuter}(2003)}]{Wuchterl2003}
{Wuchterl}, G. \& {Tscharnuter}, W.~M. 2003, \aap, 398, 1081

\bibitem[{Yan {et~al.}(2017)Yan, Jerabkova, \& Kroupa}]{Yan2017}
Yan, Z., Jerabkova, T., \& Kroupa, P. 2017, \aap, 607, A126

\bibitem[{{Zhang} {et~al.}(2001){Zhang}, {Fall}, \& {Whitmore}}]{Zhang2001}
{Zhang}, Q., {Fall}, S.~M., \& {Whitmore}, B.~C. 2001, \apj, 561, 727

\bibitem[{{Zhang} {et~al.}(2018){Zhang}, {Romano}, {Ivison}, {Papadopoulos}, \&
  {Matteucci}}]{Zhang2018}
{Zhang}, Z.-Y., {Romano}, D., {Ivison}, R.~J., {Papadopoulos}, P.~P., \&
  {Matteucci}, F. 2018, \nat, 558, 260

\bibitem[{{Zonoozi} {et~al.}(2016){Zonoozi}, {Haghi}, \&
  {Kroupa}}]{Zonoozi2016}
{Zonoozi}, A.~H., {Haghi}, H., \& {Kroupa}, P. 2016, \apj, 826, 89

\end{thebibliography}

\end{document}